\documentclass[aps,onecolumn,groupedaddress,superscriptaddress,showpacs,preprint]{revtex4-1}

\usepackage{graphicx}
\usepackage{color}
\usepackage{amsmath}
\usepackage{amsfonts}
\usepackage{amssymb}
\usepackage{dcolumn}
\usepackage{hyperref}
\usepackage{enumerate}
\usepackage{bbold}
\usepackage{cancel}
\usepackage{mathtools}
\usepackage{colonequals}
\usepackage[final]{changes}
\begin{document}
  \title{Competing Activists--Political Polarization}
\author{Lucas B\"ottcher}
\email{lucasb@ethz.ch}
\affiliation{Institute for Theoretical Physics, ETH Zurich, 8093 Zurich, Switzerland}
\affiliation{Center of Economic Research, ETH Zurich and CEPR, 8092 Zurich, Switzerland}
\author{Pedro Montealegre}
\author{Eric Goles}
\affiliation{Facultad de Ingenier\'ia y Ciencias, Universidad Adolfo Ib\'{a}\~{n}ez, Santiago, Chile}
\author{Hans Gersbach}
\affiliation{Center of Economic Research, ETH Zurich, 8092 Zurich, Switzerland}

\date{\today}
\begin{abstract} 
Recent empirical findings suggest that societies have become more polarized in various countries. That is, the median voter of today represents a smaller fraction of society compared to two decades ago and yet, the mechanisms underlying this phenomenon are not fully understood. Since interactions between influential actors (``activists'') and voters play a major role in opinion formation, e.g.~through social media, we develop a macroscopic opinion model in which competing activists spread their political ideas in specific groups of society. These ideas spread further to other groups in declining strength. While unilateral spreading shifts the opinion distribution, competition of activists leads to additional phenomena: Small heterogeneities among competing activists cause them to target different groups in society, which amplifies polarization. For moderate heterogeneities, we obtain target cycles and further amplification of polarization. In such cycles, the stronger activist differentiates himself from the weaker one, while the latter aims to imitate the stronger activist.
\end{abstract}
\maketitle
\section{Introduction}
\label{sec:intro}
There is a well-documented increase of political polarization in the US and, to a lower extent, in Europe. In the US, both the electorate and the political parties have become increasingly polarized (see e.g.~Refs.~\cite{gentzkow2016polarization,shapiro17,prior13,mccarty2016polarized,PewPolarization}).
Recent findings suggest that these developments can be traced back to a growing divide between political and cultural groups, and a lower heterogeneity within groups~\cite{gentzkow2016polarization,desmet2018cultural}. That is, we observe a simultaneous increase of polarization of parties and the electorate, and an increase in party identification and attachment to views on certain political issues. Loosely speaking, in the past few years, it has become less likely to meet liberal republicans or conservative democrats~\cite{gentzkow2016polarization,PewPolarization}.
Many explanations for these observations have been put forward. Examples include the growing influence of media~\cite{prior13,NBERw24462}, macroeconomic developments such as growing income inequality and changes in international trade, elite polarization~\cite{Fiorina2008}, or demographic changes.

This raises the issue whether polarization is due to exogenous factors, or whether it is an endogenous phenomenon resulting from forces within the opinion formation process occurring in politically and economically stable environments. Simple explanations such as the growing use of the Internet appear insufficient to describe the observed polarization behavior. A recent analysis of the influence of the Internet on political polarization of US adults revealed that the growth in polarization is greatest in demographic groups with the lowest Internet and social media use~\cite{shapiro17}. Hence, several processes may contribute to political polarization. Mathematical models can provide important insights into the dynamics of opinion formation, polarization dynamics, and related spreading processes~\cite{del2017modeling,krueger17,chuang2018age,boettcher14,boettcher16,bottcher2017targeted,bottcher2018dynamical,bottcher2018clout,boettcher171,boettcher162,bottcher2017temporal}.  In particular, agent-based and network approaches helped to identify mechanisms underlying consensus, polarization, and fragmentation dynamics~\cite{sznajd2000opinion,deffuant2000mixing,Hegselmann2002Opinion,lorenz2006consensus,galam2007role,lorenz2007continuous,Redner2019Reality}.

We propose an analytically accessible model of political change that is based on three  processes: (i) emergence of political (or cultural) innovations or ideas, (ii) dynamic diffusion of innovations across the electorate, and (iii) injection of ideas by competing influential social actors at appropriate places in society. Usually, influential actors can be individuals or groups of individuals with a particular political or cultural interest. Often, leaders of interest groups or political parties are the influential actors. Henceforth, we simply refer to such influential actors as ``activists''. Today, activists have enhanced means to spread their ideas in particular subgroups of society. Facebook and Twitter are prominent examples of social media channels that allow activists to spread their ideas in their follower groups. Recent developments in data-driven campaigning and microtargeting~\cite{kreiss18,bennett16,persily17,cadwalladr17,hoferer2019impact} provide activists with opportunities to localize and target certain voter groups by linking different data sets.
For example, in the campaigns of Ted Cruz, Donald J.~Trump, and the pro-Brexit Vote Leave movement, Facebook-user-data was used by Cambridge Analytica~\cite{Kelly2018TedCruz} and AggregateIQ~\cite{Guardian2018Brexit} to microtarget entire voter groups.

With our model, we aim at examining how a society may become polarized, or more polarized, when activists try to impact the opinions of citizens with new political ideas in order to maximize their share of supporting individuals. We consider a society in which opinion formation without political innovations and activists tends to lead to non-polarized politics in the sense that opinions of citizens are uniformly distributed across the opinion spectrum, or centered around the median. This allows us to isolate the role of competing activists that inject political ideas at certain locations in the political opinion spectrum. Our approach complements existing studies~\cite{borghesi06,vicario16,krueger17,chuang2018age} by considering the dynamical interactions of voters and activists. We demonstrate that polarization emerges in the presence of competing activists, their mutual positioning in society, and the dynamic diffusion of ideas in society. 
\section{The model}
\begin{figure}
\centering
\includegraphics[width=0.49\textwidth]{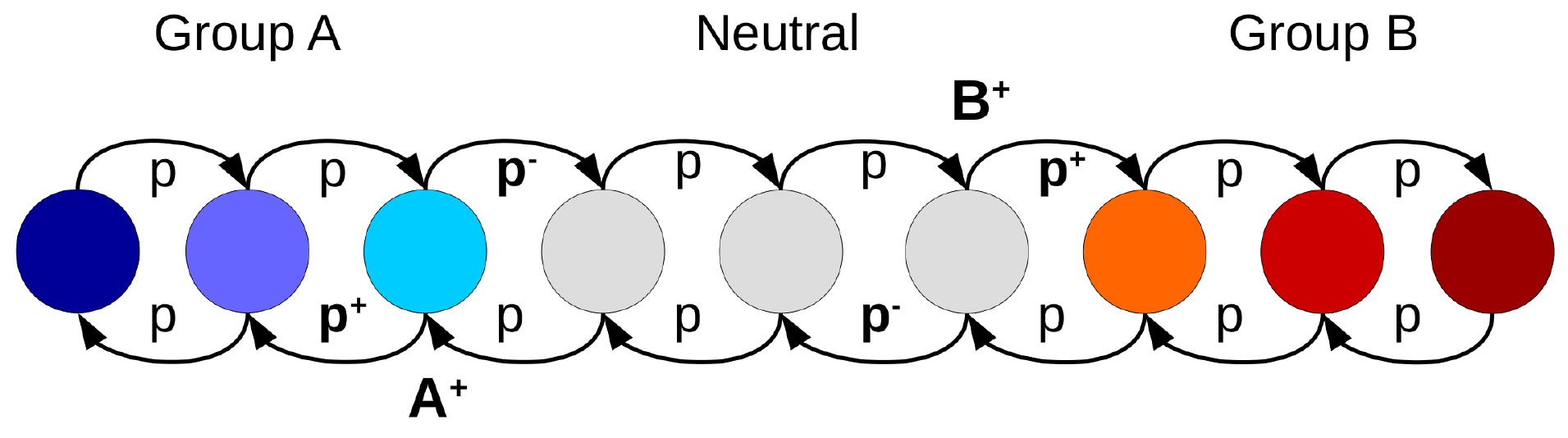}
  \caption{\textbf{Polarization model.} In this example, the political spectrum consists of $N=9$ different states and is divided in three groups: group $A$, a neutral set of agents, and group $B$. A transition from one state to its nearest neighbors occurs with probability $p$. A political activist $A^+$ or $B^+$ can locally decrease transition probabilities ($p^-<p$) or increase them  ($p^+>p$).} 
 \label{fig:model1}
\end{figure}
\begin{figure}

	\begin{minipage}{0.49\textwidth}
		\centering
		\includegraphics[scale=.9]{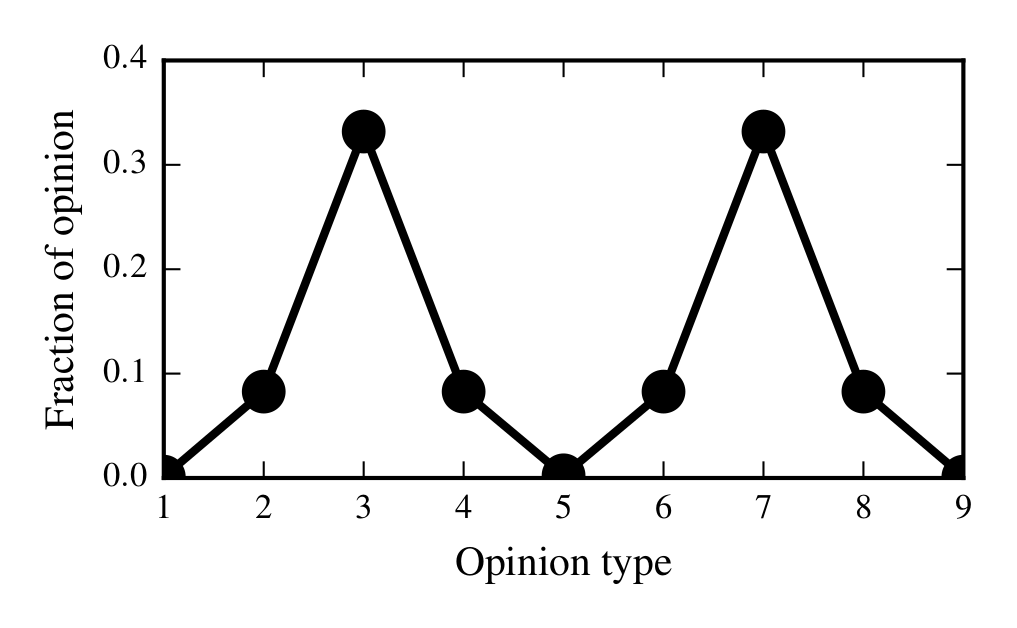}
	\end{minipage}
	\begin{minipage}{0.49\textwidth}
		\centering
		\includegraphics[scale=.9]{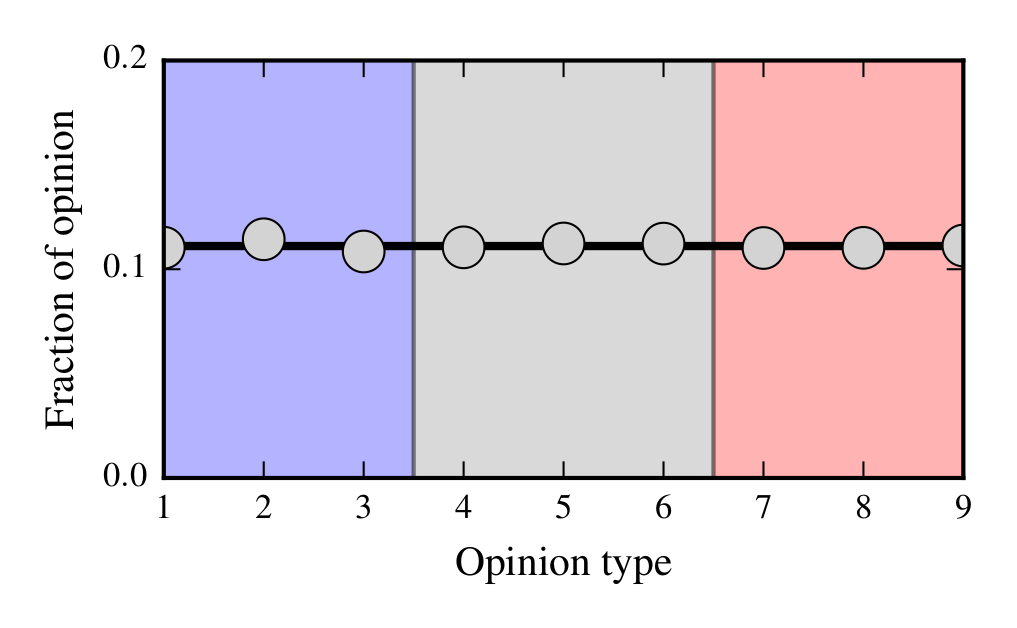}
	\end{minipage}
	  \caption{
	  	\textbf{The influence of initial distributions.}
	  	The left panel shows a polarized initial voter distribution. We observe an equilibration towards a unique uniform distribution, as illustrated in the right panel. In the right panel, numerical and analytical results (Eq.~\eqref{eq:solution_product}) are represented by grey dots and a black solid line, respectively. The data has been averaged over $10^3$ samples for $p=0.2$ and $T=2\times 10^6$ and $10^5$ initial equilibration steps.
  	 } 
	 \label{fig:no_activists}
\end{figure}
To study the influence of competing activists on opinion formation and polarization in a society, we proceed in two steps to account for processes (i--iii). In the first step, we consider a one-dimensional chain which consists of $N$ different states $\{X_i\}_{i\in\{1,\dots,N\}}$, as shown in Fig.~\ref{fig:model1} without considering activists. The choice of discrete opinion states $X_i$ is motivated by the possibility of activists to directly influence large subgroups of the society, with the transmission of a certain idea. The discretization matches the size of these subgroups. Also, the empirical opinion distributions in the US public and congress that are based on survey data~\cite{gentzkow2016polarization,PewPolarization,lewis2018} and corresponding scaling methods (e.g., the NOMINATE method~\cite{poole1985spatial,poole1984polarization}) are assembled in discrete form.
The state $X_i(t)$ represents the fraction of individuals of type $i$ at time $t$ which are normalized according to 
\begin{equation}
\sum_{i=1}^{N} X_i(t) = 1.
\label{eq:normalization}
\end{equation}
The transition probability from state $i$ to state $j$ is represented by $p_{ij}$. In Fig.~\ref{fig:model1}, we only considered transitions to nearest neighbors. The elements $p_{i j}$ form the transition matrix $P$ and satisfy $\sum_{j=1}^N p_{i j}=1$. While we employ a macroscopic model with a set of opinion classes in this work, there are convenient ways to microfound the aggregate opinion formation process at the individual level. Two ways for such microfoundations are conceivable. First, the model of DeGroot~\cite{degroot1974reaching} allows to interpret transitions from one state to another as a social learning process in a group of communicating individuals. Second, the macroscopic distributions of opinions can be recovered in random matching models in which individuals have the highest chance to meet other individuals of similar opinion. After every meeting, individuals update their opinion and may switch to the opinion of their partner with some probability. Equivalently, individuals change their opinion if they meet a sufficient number of people with alternative opinions~(see e.g.~Refs.~\cite{sznajd2000opinion,bottcher2018clout,boettcher171}).
We consider the situation where $N=9$ and partition the states in three opinion groups: group $A$, a set of neutral agents, and group $B$. Our specific choice of the number of states is not affecting the results in Secs.~\ref{sec:influence} and \ref{sec:competition}. We discuss the case of an even number of states in Appendix \ref{app:even}.
An individual located at the beginning of the chain ($i=1$) can be interpreted as a very liberal democrat, whereas the end of the chain ($i=N$) corresponds to a strongly conservative republican.

To interpret polarization according to empirical survey data of Ref.~\cite{PewPolarization}, it is important to stress again that extreme positions in the opinion chain correspond to large correlations between party identification and views on certain political issues~\cite{gentzkow2016polarization}. We model the emergence of political (or cultural) innovations or ideas (process (i)) by considering a certain initial distribution of $X_i(t=0)=X_i^0$. To describe the dynamic diffusion of innovations across the electorate (process (ii)), we now focus on the dynamics of the model. We simulate the time evolution of $X_i(t)$ for $T$ transitions, i.e., $t\in \{1,2,\dots,T\}$. In each round, we select a state $i$ uniformly at random from the set $\{1,\dots,N\}$. The corresponding update dynamics for all states $j\in \{1,\dots,N\}$ is
\begin{align}
X_j(t) &\rightarrow X_j(t)+p_{i j} X_i(t).
\label{eq:time_evol}
\end{align}
In the case of $j=i$, the transition probability can be also expressed as $p_{ii}=1-\sum_{i\neq j} p_{i j}$. An analytical solution of Eq.~\eqref{eq:time_evol} is presented in the following section.

In the second step, we introduce two competing activists $A^+$ and $B^+$, who aim at injecting ideas at some appropriate place in society (process (iii)). After activists have targeted a certain place, e.g.~their follower groups on Facebook or Twitter or groups they have identified through microtargeting as being potentially open to their ideas, the opinion formation process as described in Eq.~\eqref{eq:time_evol} applies. The goal of activists is to maximize the support of the groups on the left or right side of the political spectrum. Their ideas represent issues that are particularly attractive to the left or right side. Typical examples of such issues include Obamacare, an extension or abolishment of abortion rights, and commitments to never rise taxes, or the opposite. Activists have to choose a location $i$ in society where to inject their ideas. An activist $A^+$ ($B^+$) locally increases (decreases) $p_{i i-1}$ and decreases (increases) $p_{i i+1}$ by an amount of $\epsilon_A$ ($\epsilon_B$). Until Sec.~\ref{sec:competition}, we consider the case where $\epsilon=\epsilon_A=\epsilon_B$.

As an example, in Fig.~\ref{fig:model1}, activist $A^+$ is located at state $i=3$ and $B^+$ at state $i=6$. Mathematically, we assume that this leads to a larger transition probability $p_{3 2}=p^+=p+\epsilon$, whereas the transition probability $p_{3 4}=p^-=p-\epsilon$ is reduced. The parameter $\epsilon$ is taken from the interval $\epsilon \in (0,\min\{p,1-p\})$. We will explore two main variants how such activists compete for support. First, we explore the consequences when one or several activists have a fixed place in the political spectrum. Second, we consider the impact on polarization when two activists with possibly different strengths of ideas, and thus possibilities to affect transition probabilities, choose the best possible locations in the political spectrum. The strategic choice of locations defines a game, and we determine the mutual best responses.
\section{The influence of activists on opinion distributions}
\label{sec:influence}
\begin{figure}
	\centering
	\begin{minipage}{0.49\textwidth}
		\centering
		\includegraphics[scale=.9]{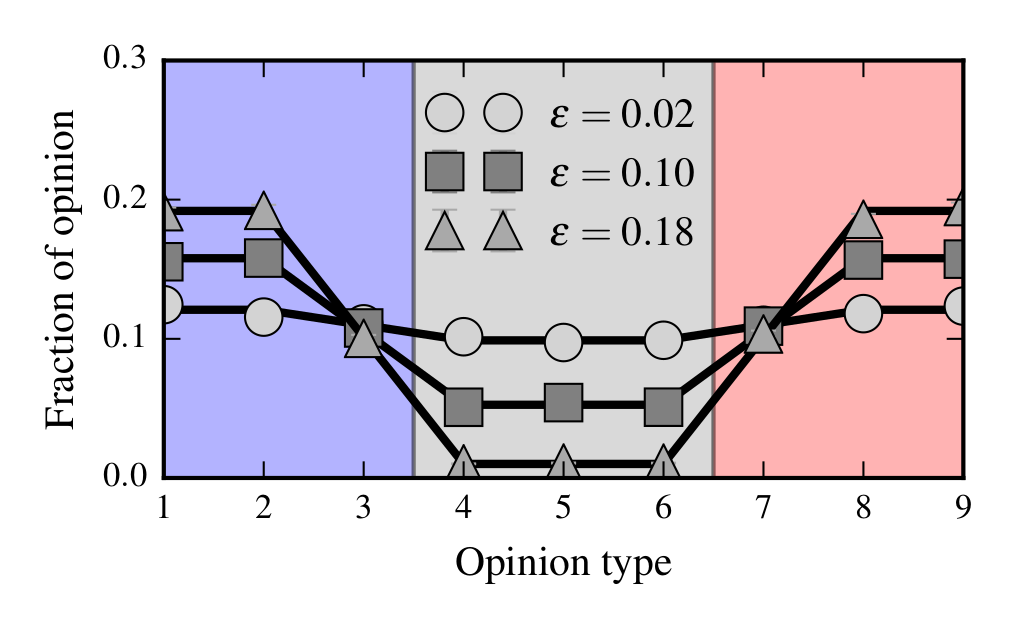}
	\end{minipage}
	\hfill
	\begin{minipage}{0.49\textwidth}
		\centering
		\includegraphics[scale=.9]{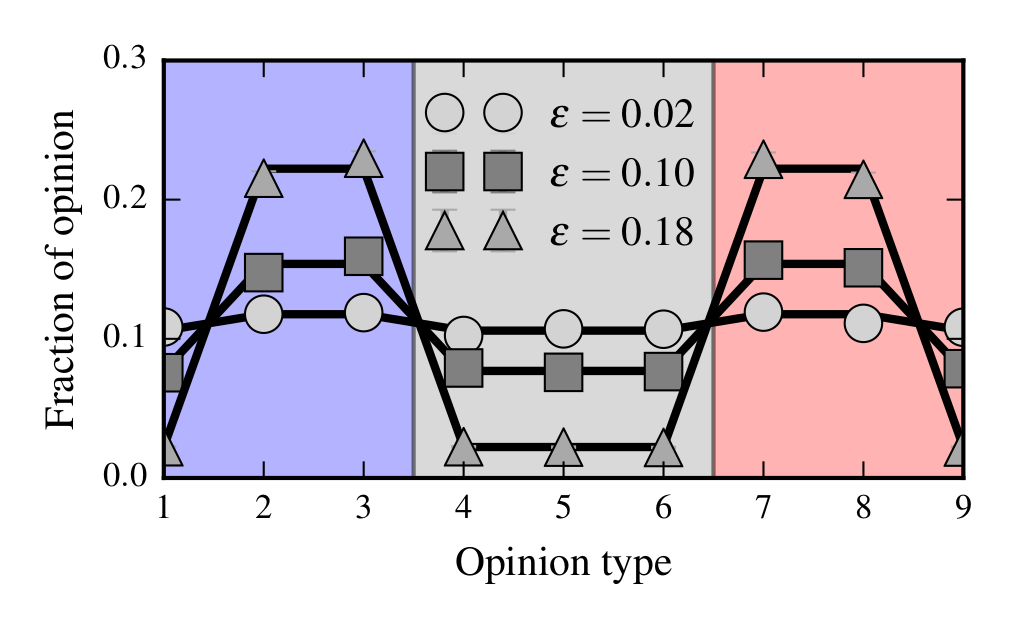}
	\end{minipage}
	  \caption{
	  	\textbf{The influence of activists.}
	  	In the left panel, activists $A^+$ and $B^+$ are located at positions $3$ and $7$, whereas in the right panel, they are located at positions $3$, $8$ and $2$, $7$. In both scenarios, a polarized equilibrium distribution is observed. Numerical and analytical results (Eq.~\eqref{eq:solution_product}) are represented by grey markers and a black solid line, respectively. The data has been averaged over $10^3$ samples for $p=0.2$, $T=2\times 10^6$ and $10^5$ initial equilibration steps.
  	 } 
	 \label{fig:activists}
\end{figure}
We first establish a mathematical framework for the update dynamics of state $i$. We consider a tridiagonal transition matrix that describes nearest neighbor interactions with $p_{i i+1}=p_i$, $p_{i i-1}=q_i$, and $p_{i i}=1-p_i-q_i$. The update rule of state $X_i(t)$ with $i\notin \{1, N\}$ reads
\begin{equation}
X_i(t+1)=(1-p_i-q_i)X_i(t)+q_{i+1} X_{i+1}(t)+p_{i-1} X_{i-1}(t).
\label{eq:update_states}
\end{equation}
We are not considering periodic boundaries, and thus find for state $i=1$ that
\begin{equation}
X_1(t+1)=(1-p_1)X_1(t)+q_2 X_{2}(t).
\label{eq:update_left}
\end{equation}
A stationary state implies $X_i(t+1)=X_i(t)\equiv X_i$ and Eq.~\eqref{eq:update_left} yields $X_{2}=\left(p_1/q_2\right) X_1$. Furthermore, based on Eq.~\eqref{eq:update_states}, we obtain
\begin{equation}
X_{i+1}=\frac{p_i+q_i}{q_{i+1}}X_i-\frac{p_{i-1}}{q_{i+1}} X_{i-1}.
\label{eq:induction1}
\end{equation}
The solution of Eq.~\eqref{eq:induction1} is given by
\begin{equation}
X_{i+1}=\left( \prod_{j=1}^i \frac{p_j}{q_{j+1}} \right) X_1\quad \text{with} \quad i\in\{1,2,\dots,N\}.
\label{eq:solution_product} 
\end{equation}
We proof this claim by induction and note that Eq.~\eqref{eq:induction1} is fulfilled for $i=1$. For the induction step, we obtain
\begin{align}
\begin{split}
X_{i+1}&=\frac{p_i+q_i}{q_{i+1}}X_i-\frac{p_{i-1}}{q_{i+1}} X_{i-1}\\
&=\frac{p_i+q_i}{q_{i+1}}\left(\frac{p_1}{q_2}\cdot \frac{p_2}{q_3}\cdots \frac{p_{i-2}}{q_{i-1} } \cdot \frac{p_{i-1}}{q_i}\right) X_1 \\
&-\frac{p_{i-1}}{q_{i+1}}\left(\frac{p_1}{q_2}\cdot \frac{p_2}{q_3}\cdots \frac{p_{i-2}}{q_{i-1}}\right) X_1\\
&=\left(\frac{p_1}{q_2}\cdot \frac{p_2}{q_3}\cdots \frac{p_{i-1}}{q_{i}} \cdot \frac{p_i}{q_{i+1}}\right) X_1=\left( \prod_{j=1}^i \frac{p_j}{q_{j+1}} \right) X_1.
\end{split}
\label{eq:induction2}
\end{align}
This proves the claim. To fulfill the normalization condition of Eq.~\eqref{eq:normalization}, we set $X_1=1$ and then divide each state $X_i$ by $\sum_{i=1}^N X_i$. The stationary distribution $X=(X_1,\dots,X_N)$ is unique because the transition matrix $P$ is irreducible and aperiodic~\cite{norris1998markov}. Irreducibility follows from the fact that any state in the Markov chain can be reached from any other state, and aperiodicity is satisfied because of $P_{ii}^n > 0$ for all $n\in\mathbb{N}$~\cite{norris1998markov}.
A different solution approach for equal transition probabilities is presented in Refs.~\cite{nagler2005directed,hauert2004dogs}.
For the example in Fig.~\ref{fig:model1}, with $p_i=q_i=p$ but without activists, the solution is $X=N^{-1}(1,1,\dots,1)^T$. We illustrate the equilibration towards the unique uniform distribution for an initially polarized distribution in Fig.~\ref{fig:no_activists}. The simulation results of Eq.~\eqref{eq:time_evol} agree well with the analytical solution given by Eq.~\eqref{eq:solution_product}.
\begin{figure}
	\centering
	\begin{minipage}{0.49\textwidth}
		\centering
		\includegraphics[scale=.9]{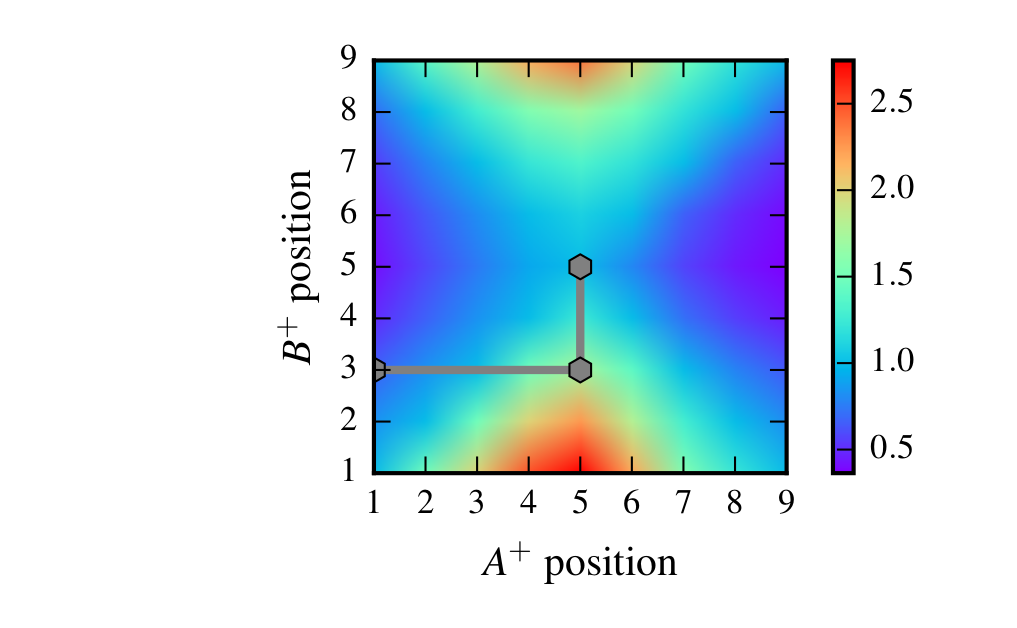}
	\end{minipage}
		\begin{minipage}{0.49\textwidth}
		\centering
		\includegraphics[scale=.9]{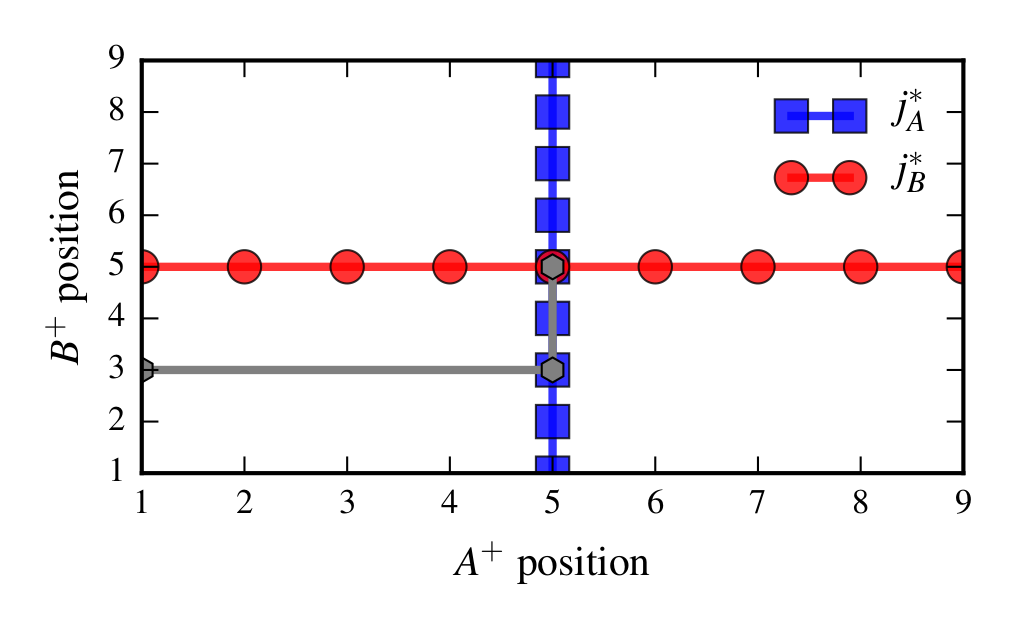}
	\end{minipage}	
	  \caption{
	  	\textbf{Relative opinion shares for different activist positions.}
	  	We compute the stationary opinion fractions according to Eq.~\eqref{eq:solution_product} for activists at different locations, which lead to locally larger ($p+\epsilon=p+0.1$) or smaller ($p-\epsilon=p-0.1$) transition probabilities ($p=0.2$). In the left panel, red indicates a large fraction of individuals in favor of group $A$ relative to group $B$, and blue indicates the opposite behavior. All states left (right) from the center are counted as belonging to group $A$ (group $B$). In the right panel, we show the best responses as defined by Eqs.~\eqref{eq:jAbest} and \eqref{eq:jBbest}. The intersection defines the corresponding Nash equilibrium. The grey solid line indicates the convergence towards the Nash equilibrium for initial activist locations $(j_A,j_B)=(1,3)$.
  	 } 
	 \label{fig:optimization1}
\end{figure}
\begin{figure*}
	\centering
	\begin{minipage}{0.49\textwidth}
		\centering
		\includegraphics[scale=.9]{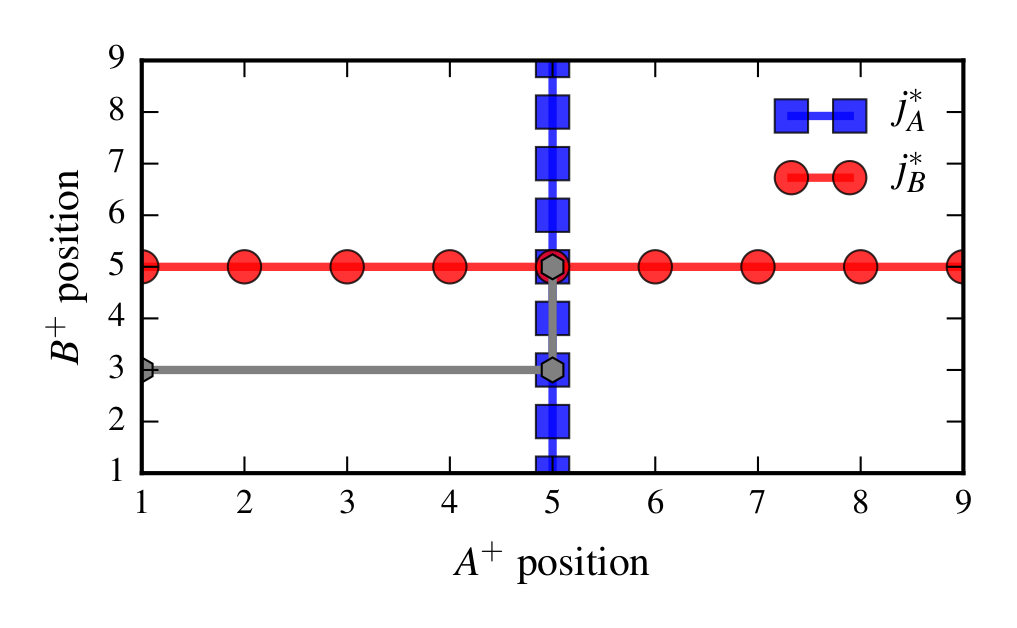}
	\end{minipage}
		\begin{minipage}{0.49\textwidth}
		\centering
		\includegraphics[scale=.9]{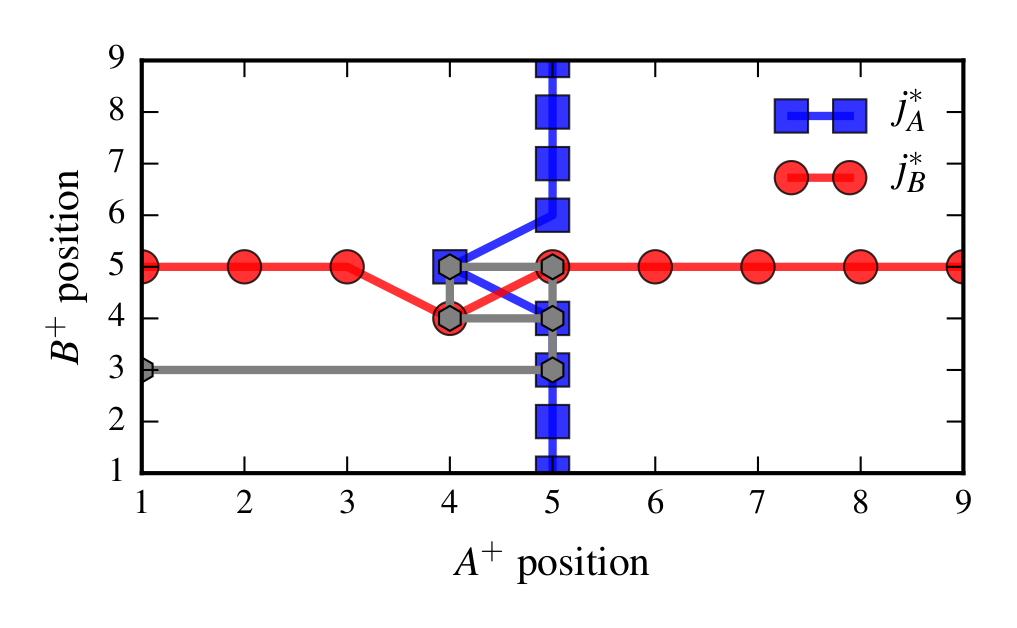}
	\end{minipage}
	\begin{minipage}{0.49\textwidth}
		\centering
		\includegraphics[scale=.9]{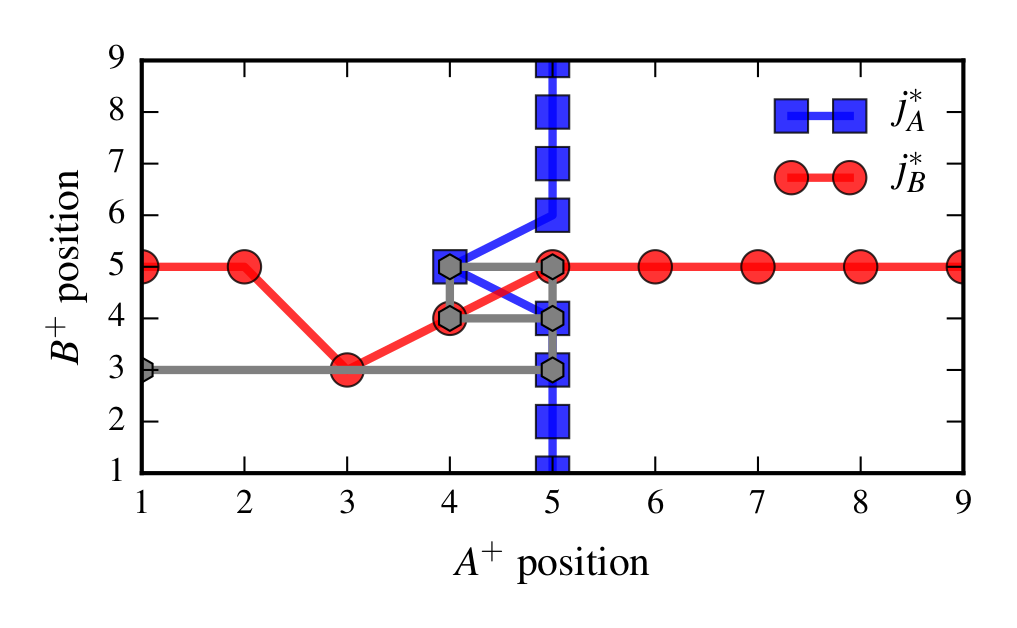}
	\end{minipage}
		\begin{minipage}{0.49\textwidth}
		\centering
		\includegraphics[scale=.9]{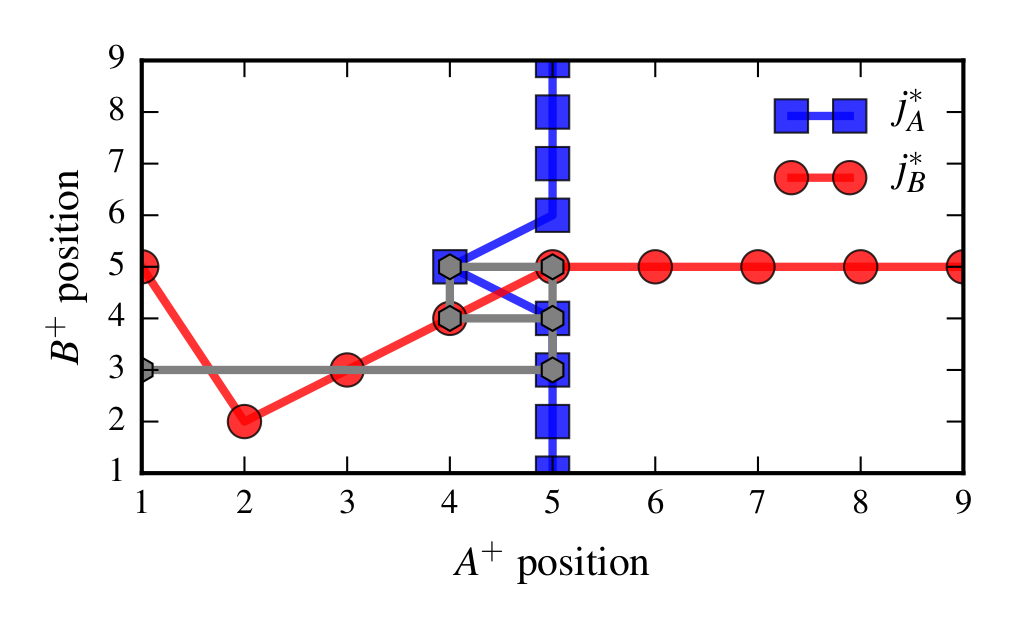}
	\end{minipage}
	  \caption{
	  	\textbf{Best responses for different activist positions and different persuasion probabilities.}
	  	Based on Eq.~\eqref{eq:solution_product}, the stationary solution is computed for activists at different locations, which lead to locally larger ($q+\epsilon_A$, $p+\epsilon_B$) or smaller ($p-\epsilon_A$, $q-\epsilon_B$) transition probabilities ($p=0.2$). From the upper left to the lower right panel, we assumed different values of $\epsilon_A \in\{0.11,0.13,0.15,0.17\}$ and set $\epsilon_B=0.1$. All states left (right) from the center are counted as belonging to group $A$ (group $B$). The grey solid line indicates the convergence towards a Nash equilibrium (upper left panel) or a cycle (remaining panels) for initial activist locations $(j_A,j_B)=(1,3)$.
  	 } 
	 \label{fig:optimization1_1}
\end{figure*}
What happens when there is one activist located at a certain position in the opinion chain? As shown in Fig.~\ref{fig:model1}, we consider an $A^+$ activist at position $i=3$ and set $q_3=p+\epsilon$, $p_{3}=p-\epsilon$. All remaining transition probabilities are unaffected and equal to $p$. The resulting unnormalized solution is $X=(1,1,p/(p+\epsilon),(p-\epsilon)/(p+\epsilon),\dots)^T$. If $\epsilon$ equals $p$,  an absorbing state emerges and the  resulting equilibrium distribution depends on the initial distribution (matrix $P$ is no longer irreducible). It is also possible to consider two or more activists which change the transition probabilities locally. We illustrate the stationary distributions for different numbers of activists and values of $\epsilon$ in Fig.~\ref{fig:activists}. Depending on the number of activists and their positions, different stationary distributions are possible. For example, in the upper panel of Fig.~\ref{fig:activists}, an $A^+$ activist targets the boundary region between group $A$ and the neutral voters, and a second $B^+$ activist targets the boundary between the neutral region and group $B$. In this way, the center is thinned out and we obtain a polarized stationary opinion distribution.
\section{Competition of two activists with endogenous location choices}
\label{sec:competition}
\begin{figure*}
	\centering
	\begin{minipage}{0.49\textwidth}
		\centering
		\includegraphics[scale=.9]{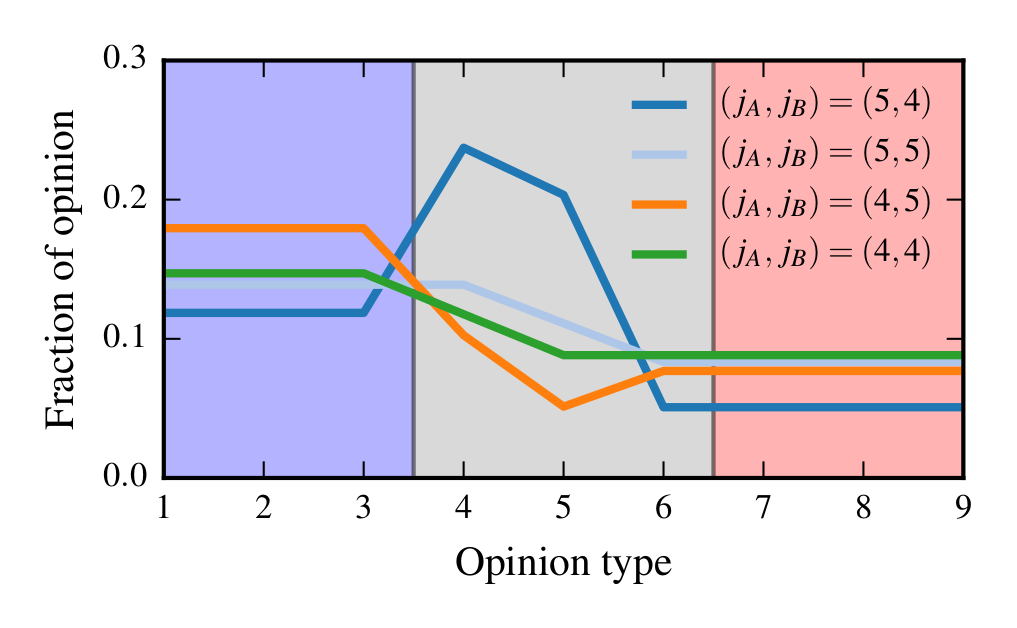}
	\end{minipage}
		\begin{minipage}{0.49\textwidth}
		\centering
		\includegraphics[scale=.9]{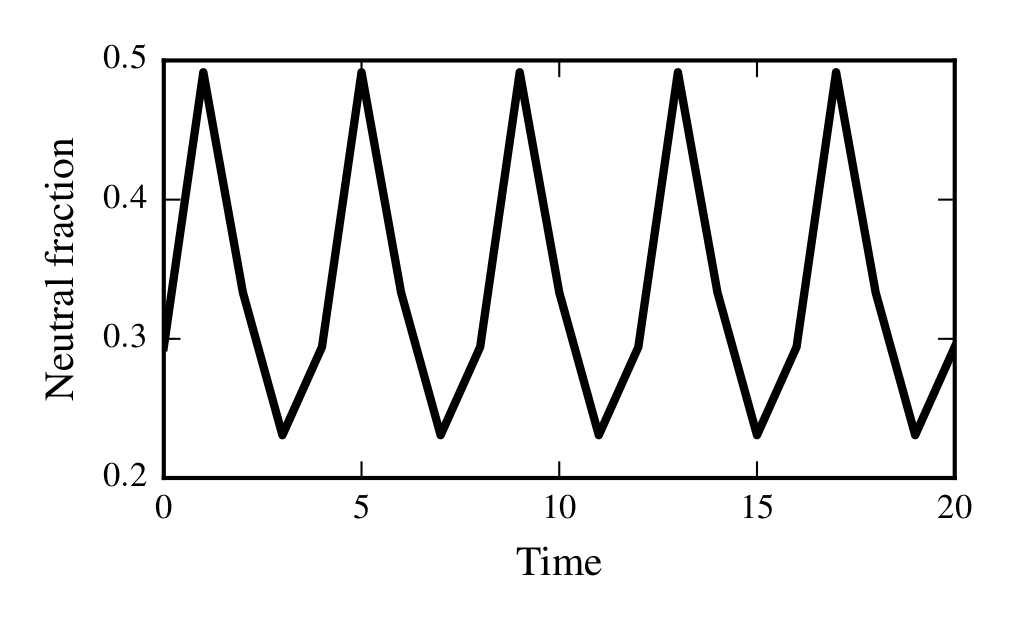}
	\end{minipage}
	\begin{minipage}{0.49\textwidth}
		\centering
		\includegraphics[scale=.9]{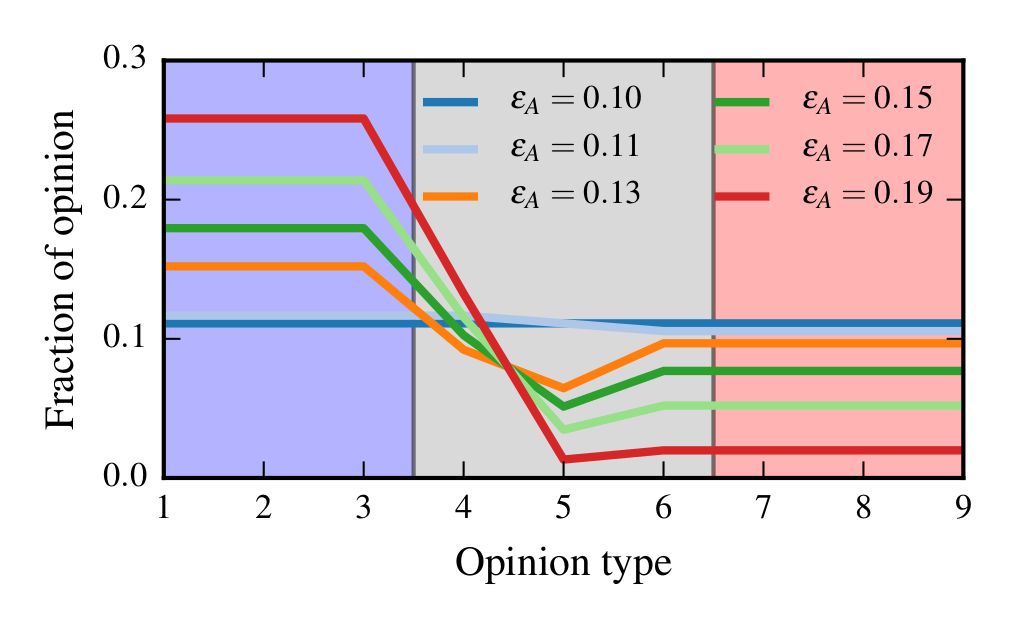}
	\end{minipage}
		\begin{minipage}{0.49\textwidth}
		\centering
		\includegraphics[scale=.9]{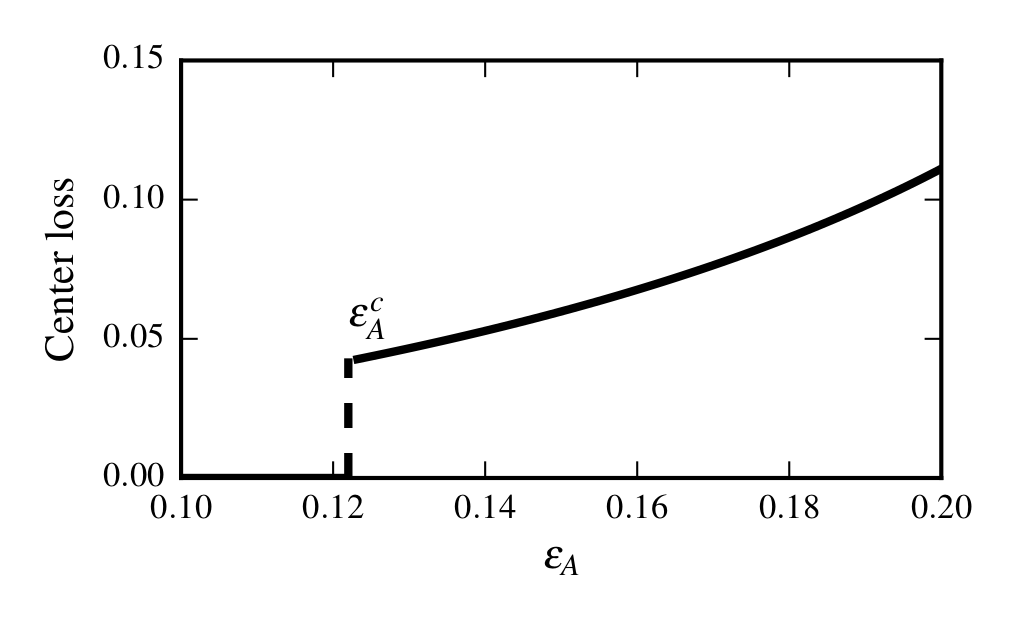}
	\end{minipage}
	  \caption{
	  	\textbf{Dynamic equilibria and threshold behavior.}
	  	Based on Eq.~\eqref{eq:solution_product}, the stationary solution is computed for activists at different locations which, lead to locally larger ($q+\epsilon_A$, $p+\epsilon_B$) or smaller ($p-\epsilon_A$, $q-\epsilon_B$) transition probabilities ($p=0.2$). In the upper panels, we set $\epsilon_A=0.15$ and $\epsilon_B=0.1$. The stationary opinion distributions of one cycle are shown in the upper left panel. In the upper right panel, we illustrate the corresponding time evolution of the opinion fraction in the neutral region. The lower left panel shows the opinion distributions for the best response of $A^+$ given $j_B=5$ and different values of $\epsilon_A$ and $\epsilon_B=0.1$. In the lower right panel, we show the center loss, i.e., the difference between the fractions at the center for $0.1\leq \epsilon_A \leq 0.2$ and $\epsilon_B=0.1$.
  	 } 
	 \label{fig:optimization1_1_distributions}
\end{figure*}
We now turn to our main analysis and consider the case, when an activist can change their position in society in response to the activity of a second activist. We perform this exercise in an initially unpolarized society (i.e., a uniform opinion distribution). This may represent the situation in the US some decades ago, when the society was much less polarized. However, apart from this empirical rationale, there are also two conceptual arguments why starting with an unpolarized society is useful. First, the enhanced means of targeting subgroups developed in recent times also allow activists to switch easily from one group to other groups over time. Hence, we identify how such enhanced targeting possibilities impact polarization. Second, by considering initially polarized societies in Appendix~\ref{app:polarized_init}, we can identify the influence of activists when polarizing activities have already been present in the past. We interpret competition of activists as an optimization game in which they try to find the optimal position in order to maximize the shares of individuals following their opinion. Specifically, let $j_A$ and $j_B$ represent the locations of activists $A^+$ and $B^+$, and let $X_i(j_A,j_B)$ be the size of the resulting equilibrated opinion group $i\in \{1,\dots, 9\}$. For a given $j_B\in\{1,\dots,9\}$, we want to obtain the best responses $j_A^\ast$ of activist $A^+$. Therefore, we consider the case where the goal of activist $A^+$ is to obtain the greatest possible fraction of voters in the left half of the opinion chain. The resulting optimal position of $A^+$ is  
\begin{equation}
j_A^\ast= \underset{j_A}{\textrm{argmax}}\left\{\sum_{i=1}^4 X_i(j_A, j_B)\right\}.
\label{eq:jAbest}
\end{equation}
Similarly, we obtain the best responses $j_B^\ast$ of an activist $B^+$ for a given $j_A\in\{1,\dots,9\}$ according to
\begin{equation}
j_B^\ast= \underset{j_B}{\textrm{argmax}}\left\{\sum_{i=6}^9 X_i(j_A, j_B)\right\}.
\label{eq:jBbest}
\end{equation}
As an example, we analyze the situation where both activists have an equally strong influence $\epsilon=0.1$ on the transition probabilities $p=0.2$. We illustrate the fraction of individuals in favor of opinion group $A$ for different locations of $A^+$ and $B^+$ in the upper panel of Fig.~\ref{fig:optimization1}. To illustrate the emergence of the Nash equilibrium, we consider the case where
$B^+$ is initially located at state $3$. The best response of $A^+$ is to target opinion state $5$. Also activist $B^+$ is then located at position $5$ as a best response. We show the corresponding trajectory (grey solid line) in Fig.~\ref{fig:optimization1}. Indeed, in the case of an equally strong influence of both activists, the best response of both activists is to always occupy the center. This behavior is also illustrated in the lower panel of Fig.~\ref{fig:optimization1}. The Nash equilibrium $\left(j_A^\ast,j_B^\ast\right)=(5,5)$ is defined by the intersection of both best response curves $j_A^\ast(j_B)$ and $j_B^\ast(j_A)$. This equilibrium corresponds to the situation where the effects of both activists cancel out if their influence is equally strong. 

Interestingly, the situation is more complex if activists  differ in their strengths ($\epsilon_A \neq \epsilon_B$). To analyze the corresponding equilibria, we consider the case where $\epsilon_A>\epsilon_B=0.1$ without loss of generality. In Fig.~\ref{fig:optimization1_1} we show the best responses for different values of $\epsilon_A\in\{0.11,0.13,0.15,0.17\}$. If $\epsilon_A$ is larger than a certain critical value $\epsilon_A^c$, three important observations can be made: (a) The best response curves have no intersection point anymore and consequently, there is no Nash equilibrium in pure strategies. (b) If $A^+$ is located at the center or somewhere left from the center, the best response of $B^+$ is to occupy the same position as $\epsilon_A$ approaches $p$. (c) The stronger activist avoids to target the same position as the weaker one. We illustrate these observations by an example. We consider the path of best responses illustrated by the grey solid line in the upper right panel of Fig.~\ref{fig:optimization1_1} ($\epsilon_A=0.13$) and initially locate $B^+$ at position $j_B=3$. The best response of $A^+$ is $j_A^\ast(j_B=3)=5$. Then, $B^+$ chooses $j_B^\ast(j_A=5)=5$ as best response. Given $j_B = 5$, activist $A^+$ avoids its opponent and targets voters at position $j_A^\ast(j_B=5)=4$. Now, the best response of $B^+$ is to match the location of $A^+$ by selecting $j_B^\ast(j_A=4)=4$. The cycle of best responses between $4$ and $5$ would go on forever. This also illustrates that no Nash equilibrium in pure strategies exists. We thus employ a dynamic version of the location game and allow in each period that one activist chooses his location as a best response in terms of the location choice of its opponent. There is an additional motivation for this modeling choice. Activists need time to organize access to the subgroups they want to target. Hence, changing locations needs time, and in our model, it takes one period.
Furthermore, we consider the activists to alternate with their location choices. The alternative would be to consider mixed strategy equilibria, which, on average, yields qualitatively the same results.

The resulting stationary distributions for the described cycle are shown in the upper left panel of Fig.~\ref{fig:optimization1_1_distributions}. We see that the concentration of voters in the neutral region varies between large and small fractions. We interpret this behavior as a temporal variation between more and less polarized opinion distributions. The corresponding time evolution of the fraction of voters in the neutral region is shown in the upper right panel of Fig.~\ref{fig:optimization1_1_distributions}. We observe oscillations with a cycle period of four.
\begin{figure*}
	\centering
	\begin{minipage}{0.49\textwidth}
		\centering
		\includegraphics[scale=.9]{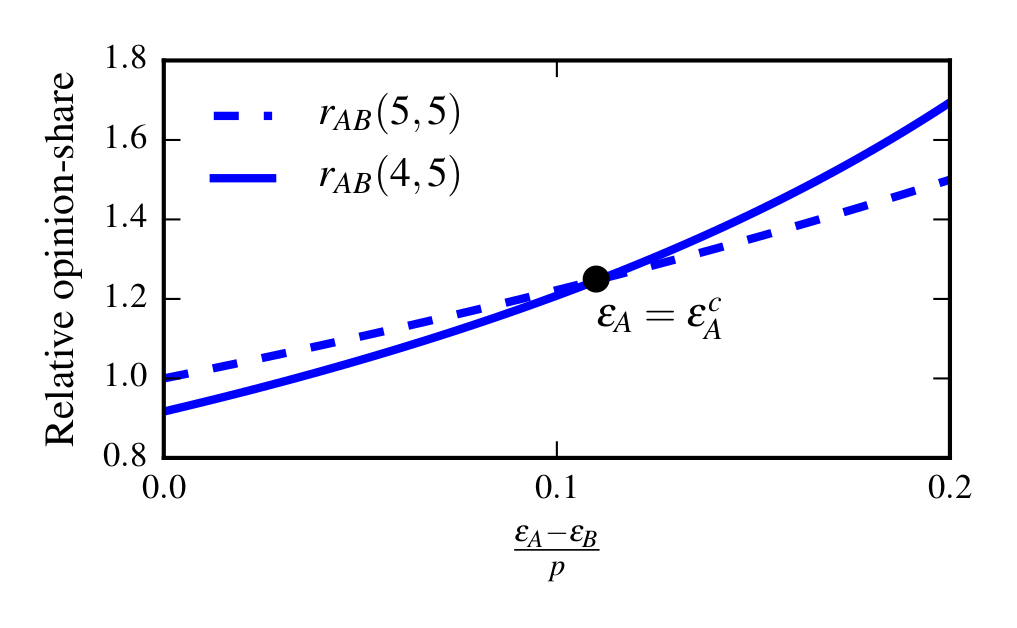}
	\end{minipage}
		\begin{minipage}{0.49\textwidth}
		\centering
		\includegraphics[scale=.9]{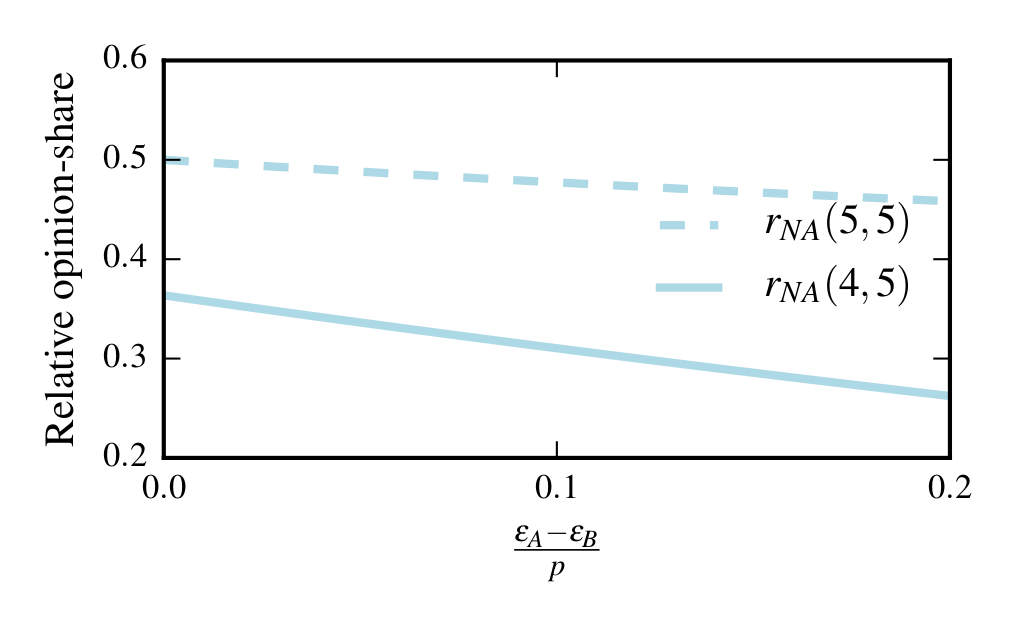}
	\end{minipage}
		\begin{minipage}{0.49\textwidth}
		\centering
		\includegraphics[scale=.9]{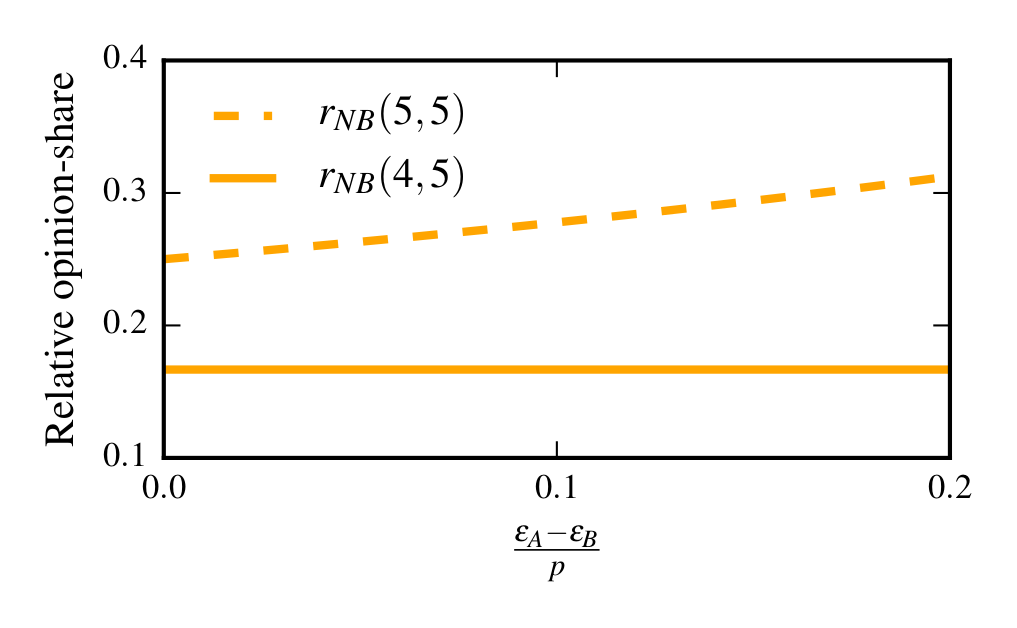}
	\end{minipage}
			\begin{minipage}{0.49\textwidth}
		\centering
		\includegraphics[scale=.9]{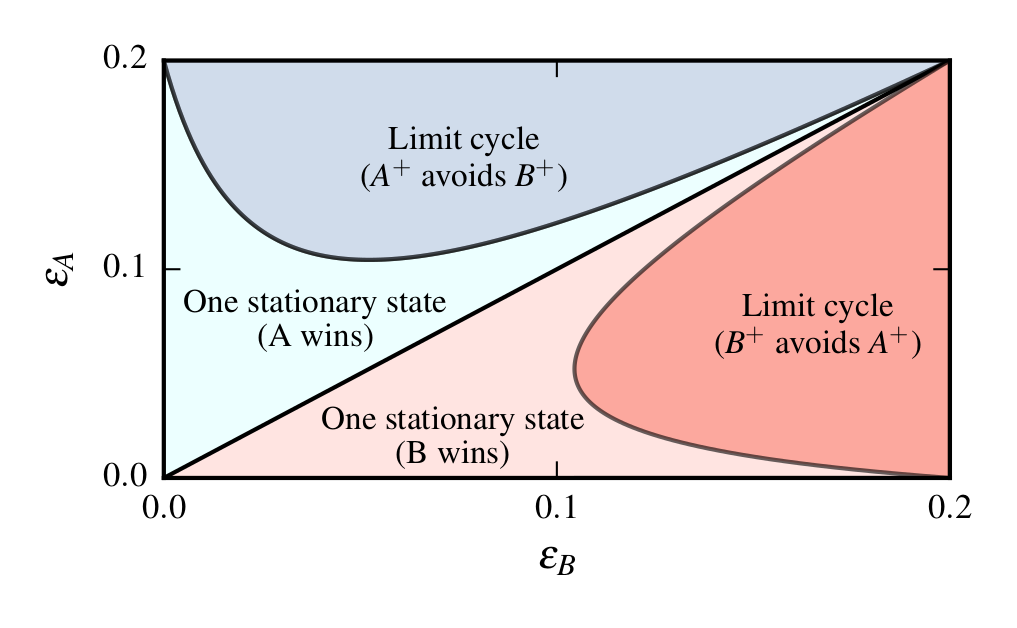}
	\end{minipage}
	  \caption{
	  	\textbf{Relative opinion-shares for certain best response activist positions.}
	  	The relative opinion-shares $r_{AB}(j_A,j_B)$ ($A$ relative to $B$), $r_{NA}(j_A,j_B)$ ($N$ relative to $A$) and $r_{NB}(j_A,j_B)$ ($N$ relative to $B$) as a function of $\left(\epsilon_A-\epsilon_B\right)/p$, with $\epsilon_B=0.1$ and $p=0.2$. The activists are either both located at the center ($(j_A,j_B)=(5,5)$) or activist $A^+$ avoids $B^+$ as a best response ($(j_A,j_B)=(4,5)$). In the case of $(j_A,j_B)=(4,5)$, the neutral regions are thinned out relative to group $A$, whereas the fraction stays constant relative to group $B$. The phase diagram of the two activist model with transition probabilities $p=0.2$ is shown in the lower right panel.} 
	 \label{fig:bestresponses}
\end{figure*}
The data presented in Fig.~\ref{fig:optimization1_1_distributions} suggests that the equilibrium where $(j_A,j_B)=(4,5)$ leads to a situation where the center is thinned out if $\epsilon_A>\epsilon_A^c$. Such a behavior is a potential amplification mechanism of polarization. To analyze this effect further, in the lower left panel of Fig.~\ref{fig:optimization1_1_distributions}, we illustrate the stationary opinion distributions, considering the best response of $A^+$, given $j_B=5$, for different values of $\epsilon_A$. If the difference between $\epsilon_A$ and $\epsilon_B$ is small enough, the neutral region is almost unaffected due to fact that the best response of $A^+$ is $j_A^\ast(j_B=5)=5$. In particular, the opinion fraction at the center is invariant under the influence of a growing $\epsilon_A$ as long as $\epsilon_A<\epsilon_A^c$. The reason for this effect is that in this case, the fraction at the center is
\begin{equation}
X_5(j_A=5,j_B=5)=\frac{\frac{p}{p+\epsilon}}{\frac{9 p}{p+\epsilon}}=\frac{1}{9}.
\end{equation}
However, as soon as $\epsilon_A$ exceeds $\epsilon_A^c$, the neutral region is suddenly thinned out and we observe a clear polarization effect. The reason is that the best response of $A^+$ is to now target position $j_A^\ast(j_B=5)=4$ as best response. The difference between the opinion fraction at the central state for $\epsilon_A=0$ and $\epsilon_A>0$ (center loss) is shown in the lower right panel of Fig.~\ref{fig:optimization1_1_distributions}. It is noteworthy that this transition only emerges due to the best response dynamics of two competing activists. Such abrupt transitions are difficult to predict and control because slight variations in the control parameter cause significant macroscopic differences.

To understand the observed threshold effect, we have to analyze the competitive advantage accompanying the evasion behavior of the stronger activist. We use Eq.~\eqref{eq:update_states} and determine the unnormalized equilibrium distribution $X=(X_1,\dots,X_9)$. We first analyze the situation in which both activists are located at the center. The resulting unnormalized distribution is $X=(1,1,1,1,p/(p+\Delta \epsilon),(p-\Delta \epsilon)/(p+\Delta \epsilon),\dots,(p-\Delta \epsilon)/(p+\Delta \epsilon))$, with $\Delta \epsilon = \epsilon_A-\epsilon_B$. In this case, the relative vote-share $r_{AB}(j_A,j_B)$ of group $A$ relative to group $B$ is given by
\begin{equation}
r_{A B}(5,5)=\frac{\sum_{i=1}^4 X_i(5,5)}{\sum_{i=6}^9 X_i(5,5)}=\frac{p+ \Delta \epsilon}{p - \Delta \epsilon}.
\end{equation}
If activist $A^+$ is, however, located at the center and activist $B^+$ is located one position further to the left, we obtain
\begin{equation}
r_{A B}(4,5)=\frac{\sum_{i=1}^4 X_i(4,5)}{\sum_{i=6}^9 X_i(4,5)}=\frac{ (4p +3 \epsilon_A)(p - \epsilon_B)}{4 (p-\epsilon_A)(p+\epsilon_B)}.
\end{equation}
These two relative opinion-shares are shown in the upper left panel of Fig.~\ref{fig:bestresponses}. The intersection point determines the threshold
\begin{equation}
\epsilon_A^c = \frac{p^2 + 7 {\epsilon_B}^2}{p + 7 \epsilon_B}.
\label{eq:critical_epsA}
\end{equation}
Furthermore, in Fig.~\ref{fig:bestresponses}, we also show the remaining relative opinion shares
\begin{equation}
r_{N A}(j_A,j_B)=\frac{X_5(4,5)}{\sum_{i=1}^4 X_i(4,5)}
\end{equation}
and
\begin{equation}
r_{N B}(j_A,j_B)=\frac{X_5(4,5)}{\sum_{i=6}^9 X_i(4,5)}.
\end{equation}
In the case of $p=0.2$ and $\epsilon_B=0.1$, the threshold $\epsilon_A^c=0.122$ is in agreement with the observations in Fig.~\ref{fig:optimization1_1_distributions}. The lower right panel of Fig.~\ref{fig:bestresponses} shows the corresponding phase diagram. There are two phases: The first phase, in which group $A$ has the majority due to $\epsilon_A>\epsilon_B$, and the second, where the rivaling group $B$ dominates. In both regions, it is either the case that the activist dynamics leads to a stable Nash equilibrium or that a limit cycle behavior with a phase separation described by Eq.~\eqref{eq:critical_epsA} emerges. In the supplementary material, we show that the described behavior is also present in the case of an even number of states $N$. The difference is that there already exist four Nash equilibria for $\epsilon_A=\epsilon_B$, due to the absence of a unique center. Furthermore, in the supplementary material, we describe that the non-existence of Nash equilibria in pure strategies is also observable for both initially polarized and unpolarized populations.
\section{Conclusion}
We have introduced a mathematical framework to study polarization effects in a one-dimensional opinion chain. Each state in the chain corresponds to a certain party identification (e.g., democrat or republican) correlated with certain political views (e.g., liberal or conservative). Transitions are possible from every state to its neighboring states with a finite probability. We account for political activists by introducing local biases in the probability flows according to a given activist influence. Our model describes polarization as a phenomenon emerging in the presence of competing activists, their mutual positioning in society, and the dynamic diffusion of ideas in society. Possible extensions of our work include the dynamics of more than two activists, multiple opinion chains representing individual states or subgroups of a society, and activists whose influence will not vanish immediately after they have left a certain state.
\acknowledgments
We acknowledge financial support from the ETH Risk Center and from the Center for Mathematical Modeling (Santiago, Chile). PM acknowledges funding from CONICYT, PAI, Convocatoria Nacional Subvenci\'on a instalaci\'on en la academia convocatoria a\~no 2017 PAI77170068.
\appendix
\section{Initially unpolarized and polarized populations}
\label{app:polarized_init} 
\begin{figure}
	\centering
	\begin{minipage}{0.49\textwidth}
		\centering
		\includegraphics[scale=.9]{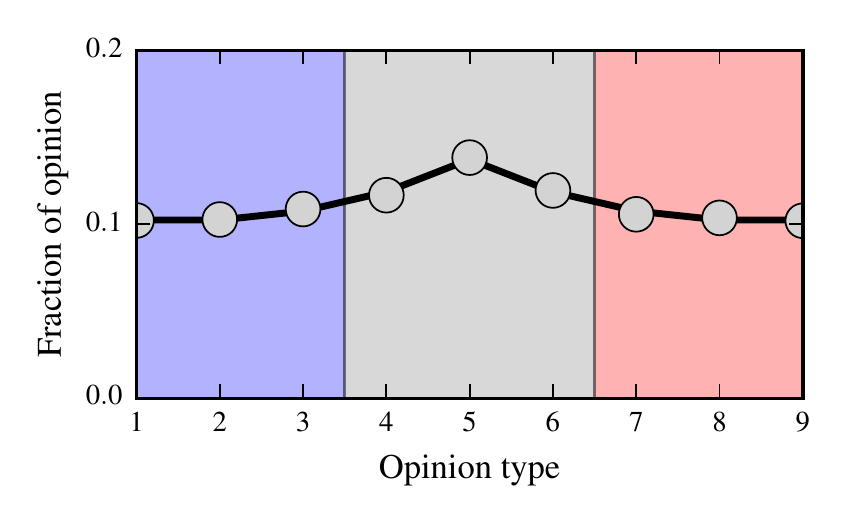}
	\end{minipage}
	\hfill
	\begin{minipage}{0.49\textwidth}
		\centering
		\includegraphics[scale=.9]{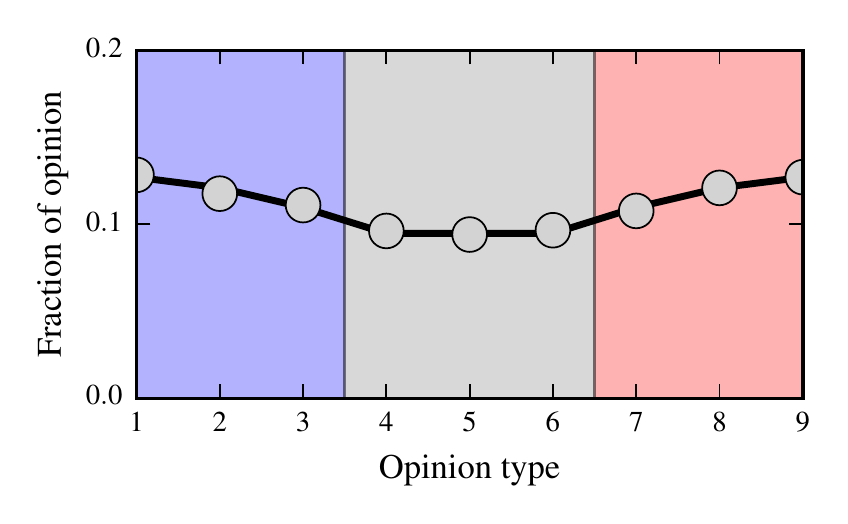}
	\end{minipage}
	  \caption{
	  	\textbf{Intrinsic anti-polarization and polarization effects.}
	  	Intrinsic anti-polarization and polarization effects are taken into account according to Eqs.~\eqref{eq:lambdaC} and \eqref{eq:lambdaE}. In the left panel, we set $\lambda_C=1.05$ and $\lambda_E=0$, whereas in the right panel, we set $\lambda_C=0$ and $\lambda_E=1.05$. Numerical and analytical results (Eq.~(6) in the main text) are represented by grey dots and a black solid line, respectively. The data has been averaged over $10^3$ samples for $p=0.2$ and $T=2\times 10^6$ and $10^5$ initial equilibration steps.
  	 } 
	 \label{fig:polarized_init}
\end{figure}

In addition to different persuasion effects of both activists $A^+$ and $B^+$ and accordingly different values of $\epsilon_A$ and $\epsilon_B$, it may also be the case that there is an intrinsic imbalance in the probability flow (i.e., an anti-polarizing or a polarizing force due to gradually different values of the transition probability $p$). We model the anti-polarization effect with a transition probability that decays according to
\begin{align}
& p_i=\lambda_C p_{i-1} \quad \text{for} \quad i\in\{2,3,4\},  \\
& q_{i-1} = \lambda_C q_i
 \quad \text{for} \quad  i\in\{9,8,7\}. 
 \label{eq:lambdaC}
\end{align}
An example of an anti-polarized stationary opinion distribution for $\lambda_C=1.05$ is shown in the left panel of Fig.~\ref{fig:polarized_init}. A value of $\lambda_C>1$ clearly leads to a larger concentration of voters at the center. In a similar way, we model a polarizing effect with 
\begin{align}
\begin{split}
&p_i =\lambda_E p_{i-1} \quad \text{for} \quad i\in\{6,7,8\},\\
&q_{i-1} = \lambda_E q_i,
 \quad \text{for} \quad  i\in\{4,3,2\}.
\end{split} 
 \label{eq:lambdaE}
\end{align}
A value of $\lambda_E>1$ leads to an initially polarized opinion distribution as shown in the right panel of Fig.~\ref{fig:polarized_init}.

\begin{figure}
	\centering
	\begin{minipage}{0.49\textwidth}
		\centering
		\includegraphics[scale=.9]{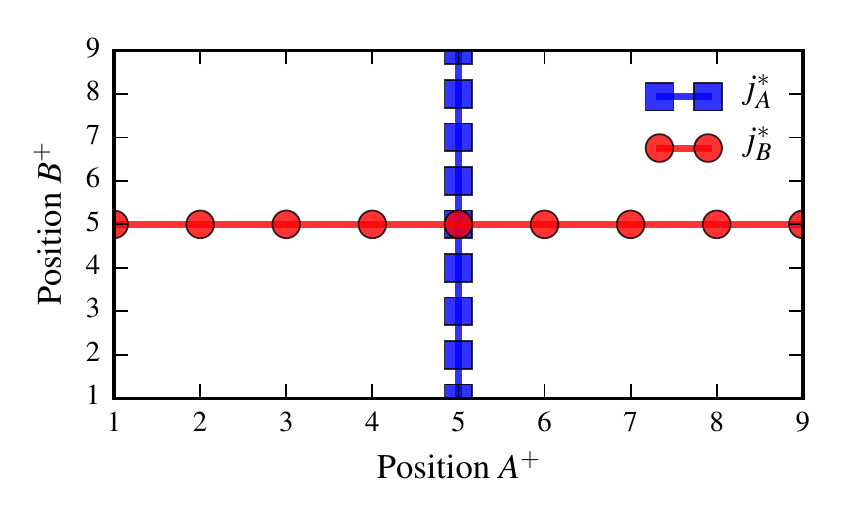}
	\end{minipage}
		\begin{minipage}{0.49\textwidth}
		\centering
		\includegraphics[scale=.9]{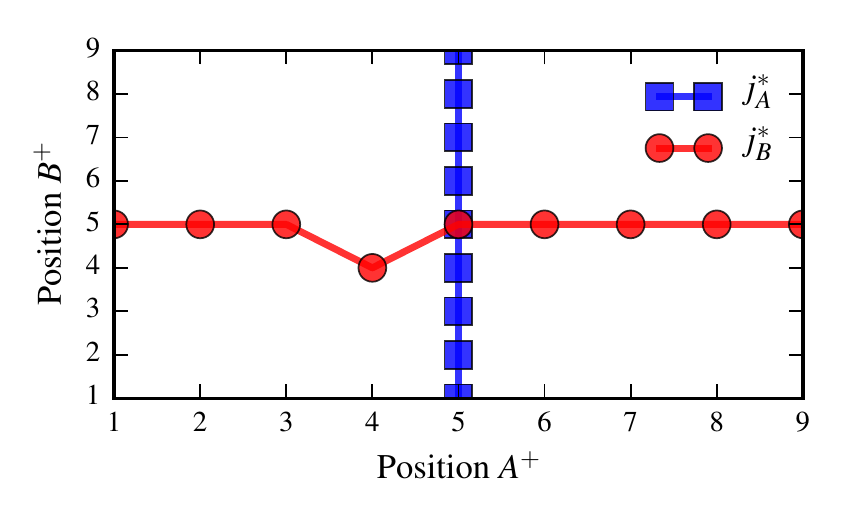}
	\end{minipage}
	\begin{minipage}{0.49\textwidth}
		\centering
		\includegraphics[scale=.9]{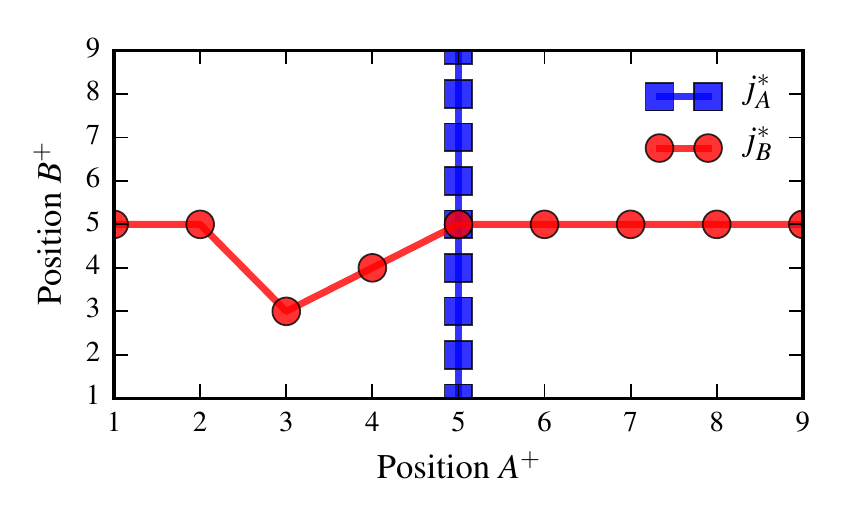}
	\end{minipage}
		\begin{minipage}{0.49\textwidth}
		\centering
		\includegraphics[scale=.9]{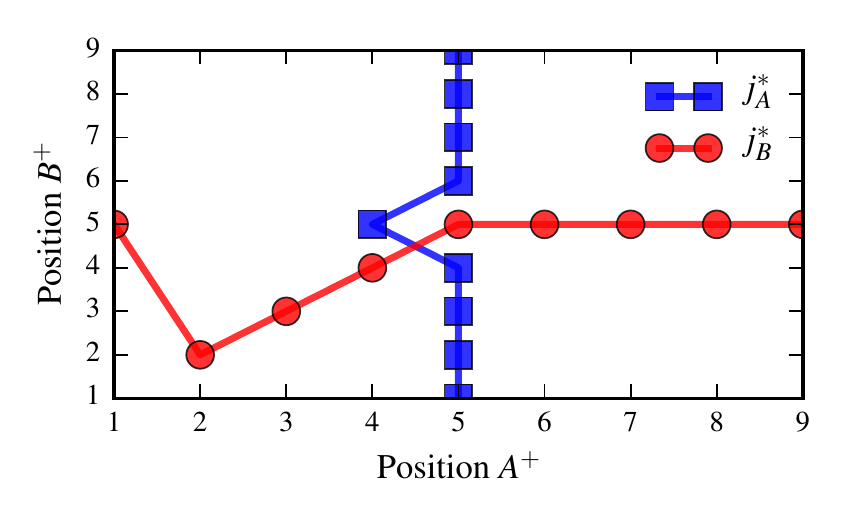}
	\end{minipage}
	\begin{minipage}{0.49\textwidth}
		\centering
		\includegraphics[scale=.9]{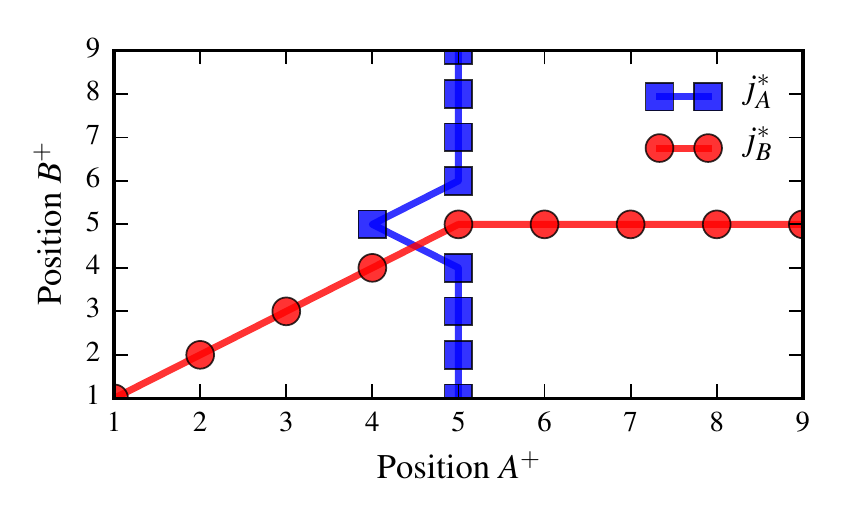}
	\end{minipage}
		\begin{minipage}{0.49\textwidth}
		\centering
		\includegraphics[scale=.9]{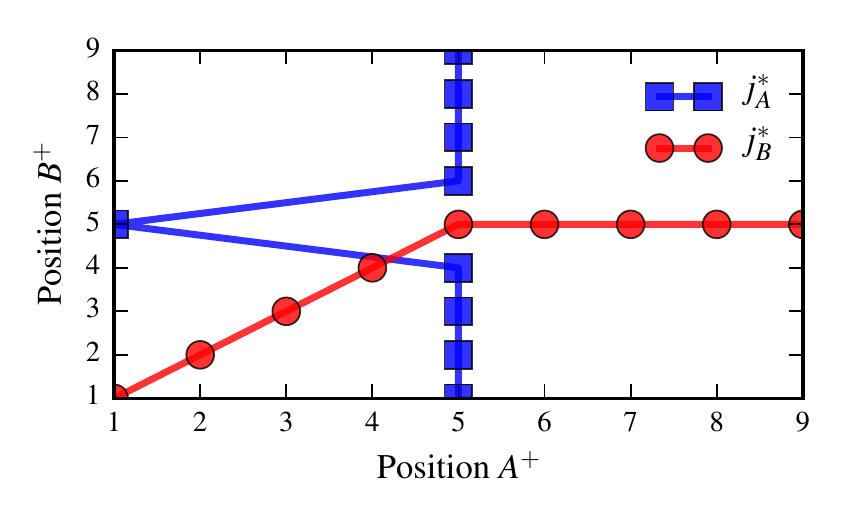}
	\end{minipage}
	  \caption{
	  	\textbf{Best responses for different activist positions and different persuasion probabilities.}
	  	Based on Eq.~(6) in the main text, the stationary solution is computed for activists at different locations, which lead to locally larger ($q+\epsilon_A$, $p+\epsilon_B$) or smaller ($p-\epsilon_A$, $q-\epsilon_B$) transition probabilities ($p=0.2$). From the upper left to the lower right panel, we assumed different values of $\epsilon_A\in\{0.15,0.16,0.17,0.18,0.19,0.199\}$ and set $\epsilon_B=0.1$. In addition, we incorporated anti-polarization effects according to Eq.~\eqref{eq:lambdaC} and set $\lambda_C=1.05$ and $\lambda_E=0$. All states left (right) from the center are counted as belonging to group $A$ (group $B$).
  	 } 
	 \label{fig:optimization1_lambdaC}
\end{figure}

\begin{figure}
	\centering
	\begin{minipage}{0.49\textwidth}
		\centering
		\includegraphics[scale=.9]{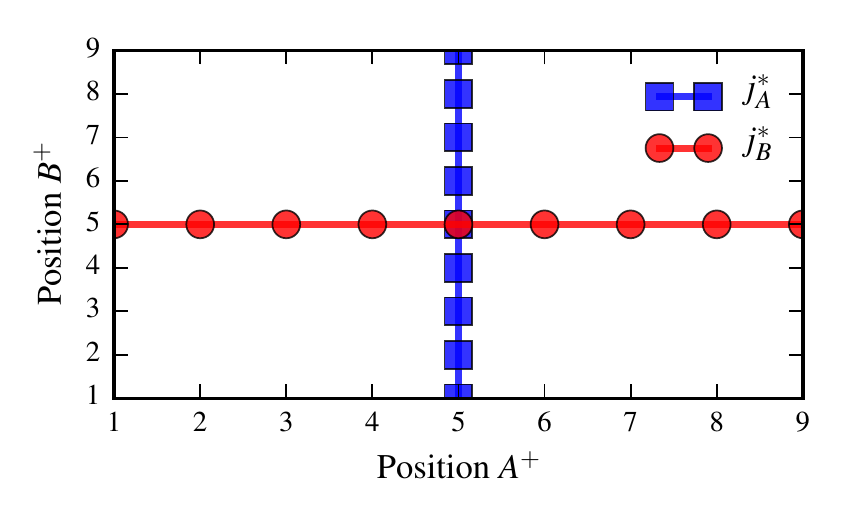}
	\end{minipage}
		\begin{minipage}{0.49\textwidth}
		\centering
		\includegraphics[scale=.9]{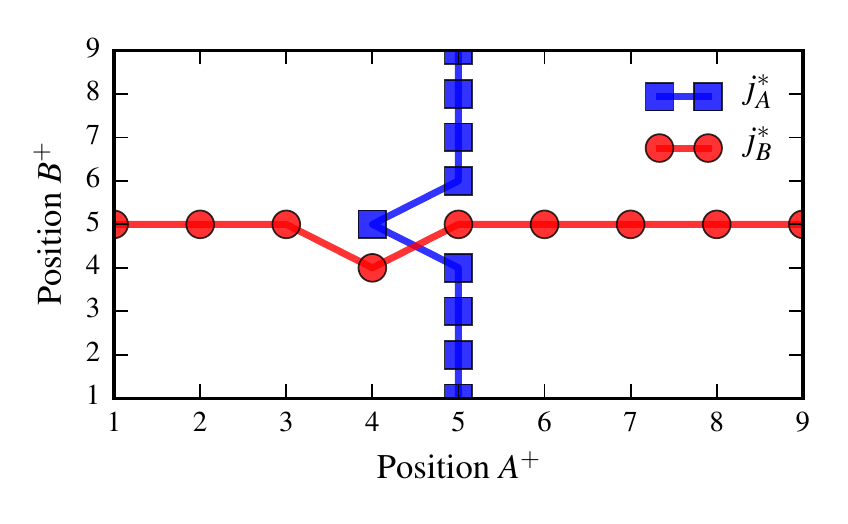}
	\end{minipage}
	\begin{minipage}{0.49\textwidth}
		\centering
		\includegraphics[scale=.9]{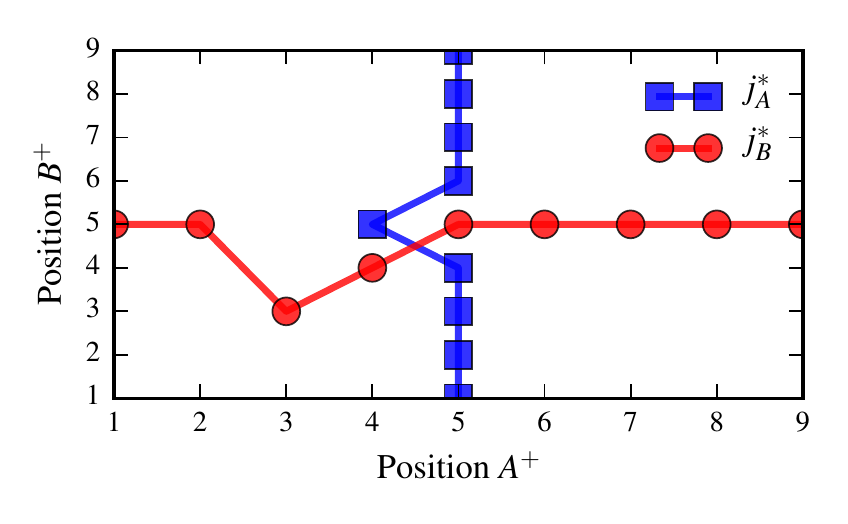}
	\end{minipage}
		\begin{minipage}{0.49\textwidth}
		\centering
		\includegraphics[scale=.9]{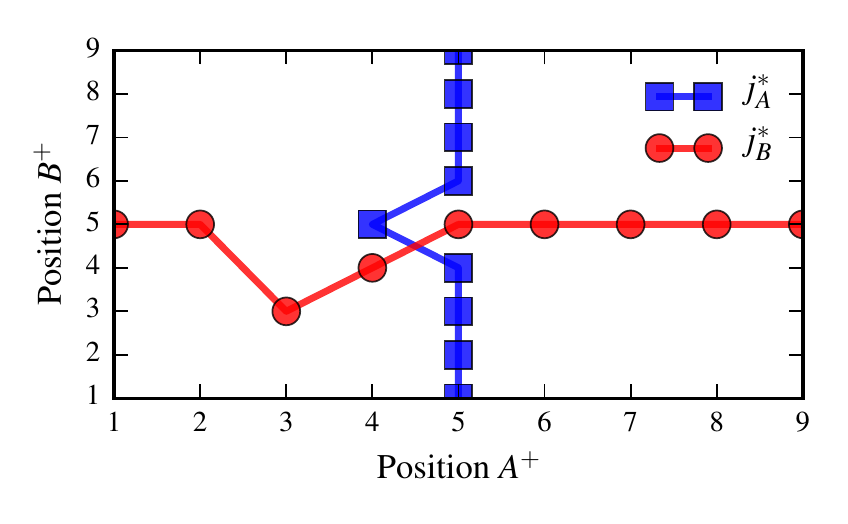}
	\end{minipage}
	\begin{minipage}{0.49\textwidth}
		\centering
		\includegraphics[scale=.9]{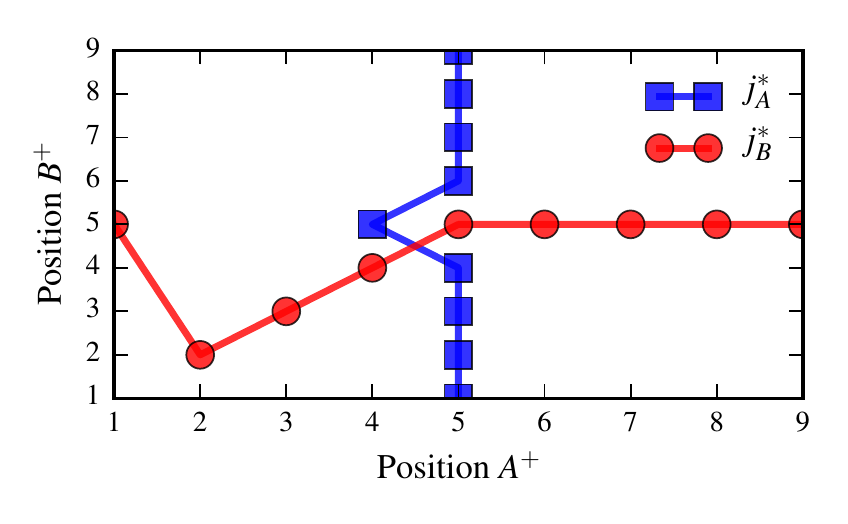}
	\end{minipage}
		\begin{minipage}{0.49\textwidth}
		\centering
		\includegraphics[scale=.9]{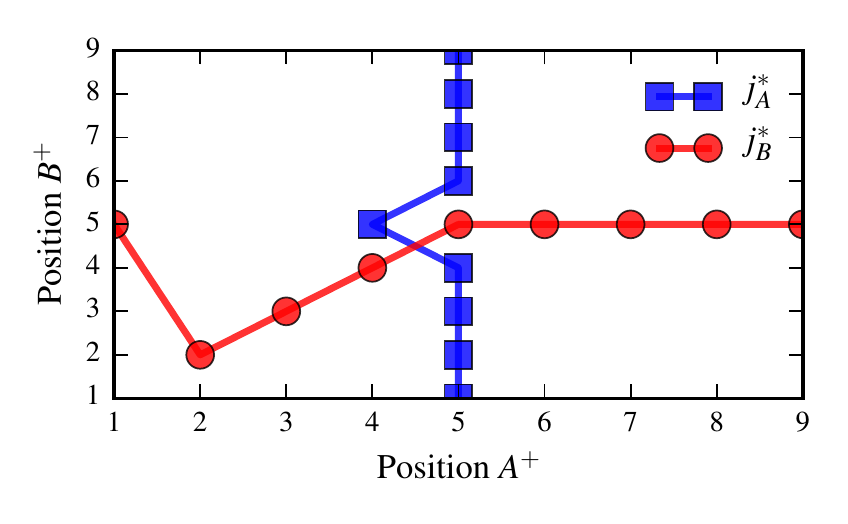}
	\end{minipage}
	\begin{minipage}{0.49\textwidth}
		\centering
		\includegraphics[scale=.9]{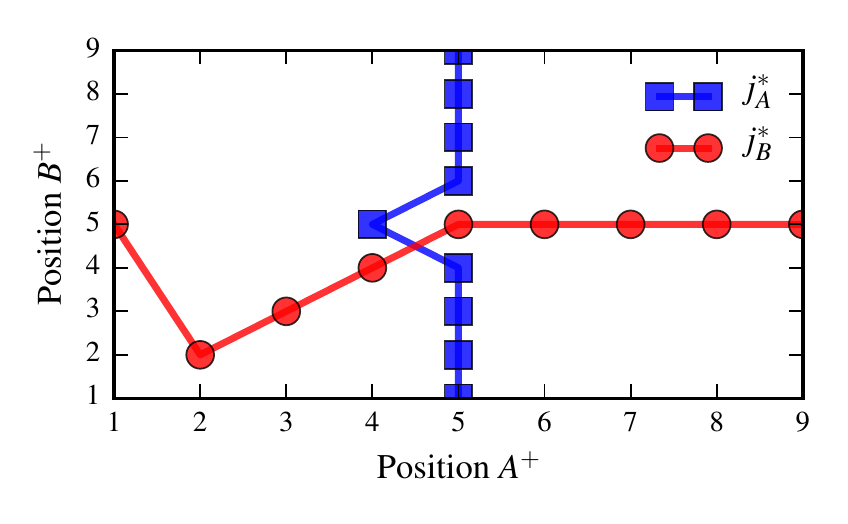}
	\end{minipage}
		\begin{minipage}{0.49\textwidth}
		\centering
		\includegraphics[scale=.9]{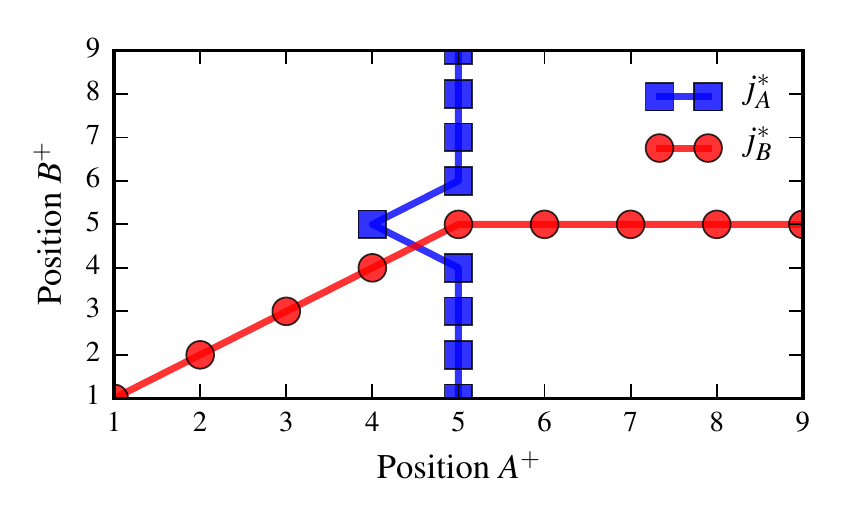}
	\end{minipage}
	  \caption{
	  	\textbf{Best responses for different activist positions and different persuasion probabilities.}
	  	Based on Eq.~(6) in the main text, the stationary solution is computed for activists at different locations, which lead to locally larger ($q+\epsilon_A$, $p+\epsilon_B$) or smaller ($p-\epsilon_A$, $q-\epsilon_B$) transition probabilities ($p=0.2$). From the upper left to the lower right panel, we assumed different values of $\epsilon\in\{0.12,0.13,\dots,0.19\}$ and set $\epsilon_B=0.1$. In addition, we incorporated polarization effects according to Eq.~\eqref{eq:lambdaC} and set $\lambda_C=0$ and $\lambda_E=1.05$. All states left (right) from the center are counted as belonging to group $A$ (group $B$).
  	 } 
	 \label{fig:optimization1_lambdaE}
\end{figure}

We also investigate the best responses for the situation where an activist $A^+$ has a larger persuasion effect according to $\epsilon_A>\epsilon_B$. We illustrate the best responses for $\lambda_C=1.05$ and $\lambda_E=0$ in Fig.~\ref{fig:optimization1_lambdaC} and for $\lambda_C=0$ and $\lambda_E=1.05$ in Fig.~\ref{fig:optimization1_lambdaE}. The observed behavior is in agreement with the one found for an initially unpolarized society in the main text. If the difference between $\epsilon_A$ and $\epsilon_B$ is small enough, a unique Nash equilibrium exists. However, if the difference becomes too large, no Nash equilibrium in pure strategies is observable anymore. Instead we find the same cycle behavior as in the main text.
\section{Even number of states}
\label{app:even}
\begin{figure}
	\centering
	\begin{minipage}{0.49\textwidth}
		\centering
		\includegraphics[scale=.9]{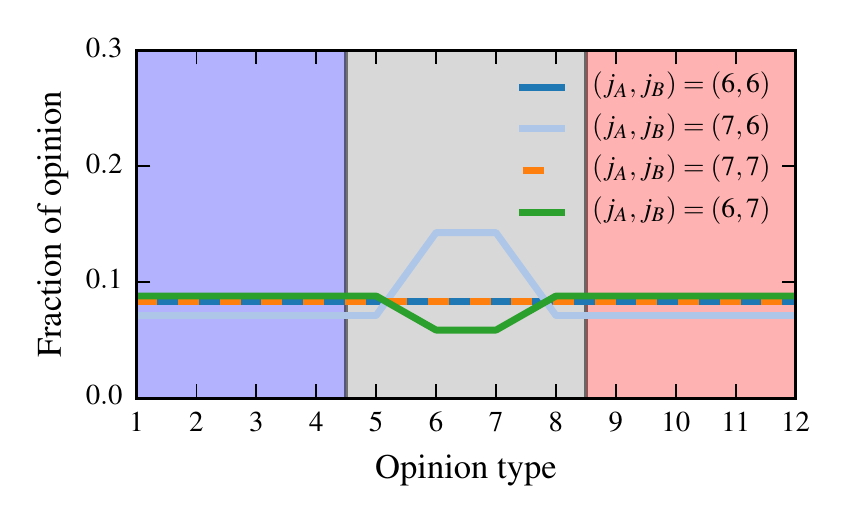}
	\end{minipage}
		\begin{minipage}{0.49\textwidth}
		\centering
		\includegraphics[scale=.9]{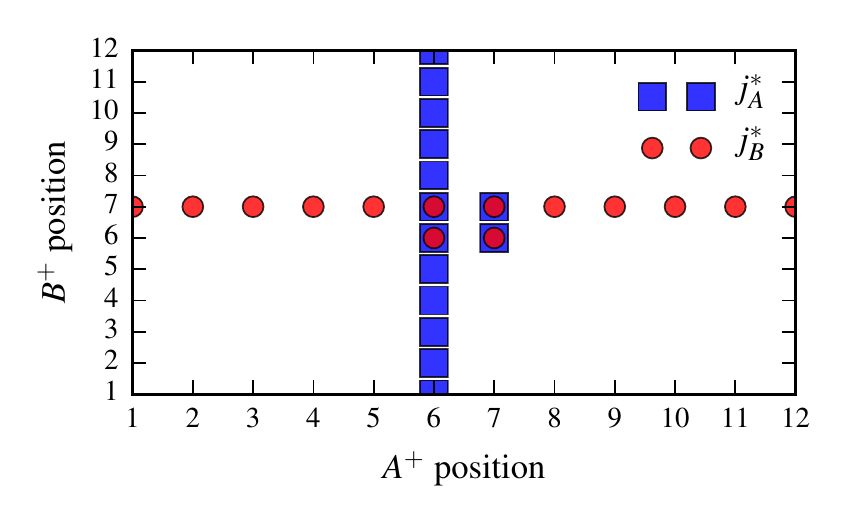}
	\end{minipage}
	\begin{minipage}{0.49\textwidth}
		\centering
		\includegraphics[scale=.9]{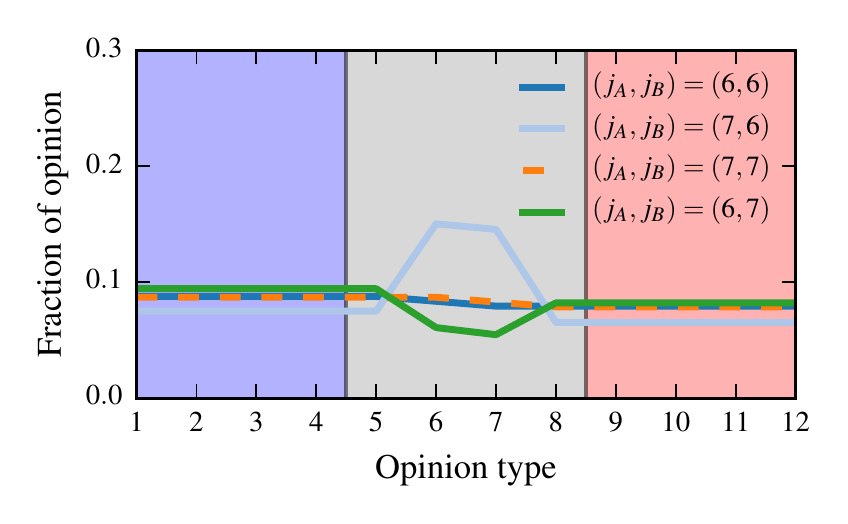}
	\end{minipage}
		\begin{minipage}{0.49\textwidth}
		\centering
		\includegraphics[scale=.9]{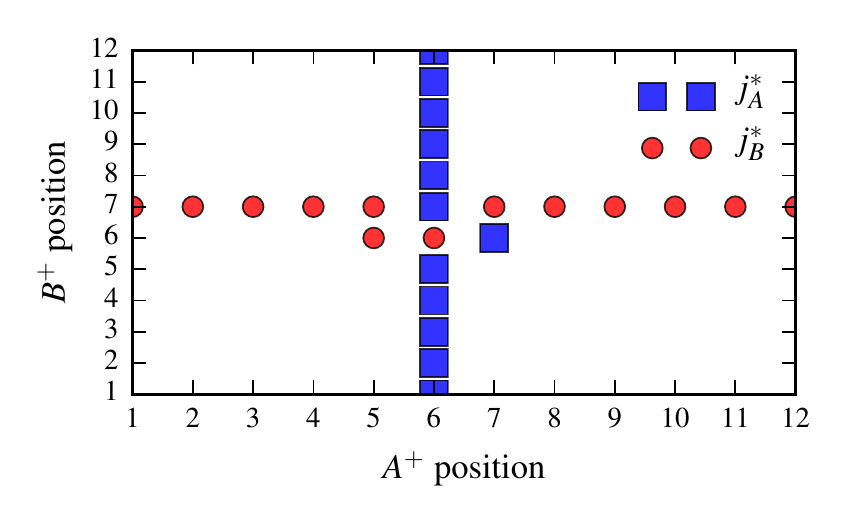}
	\end{minipage}
		\begin{minipage}{0.49\textwidth}
		\centering
		\includegraphics[scale=.9]{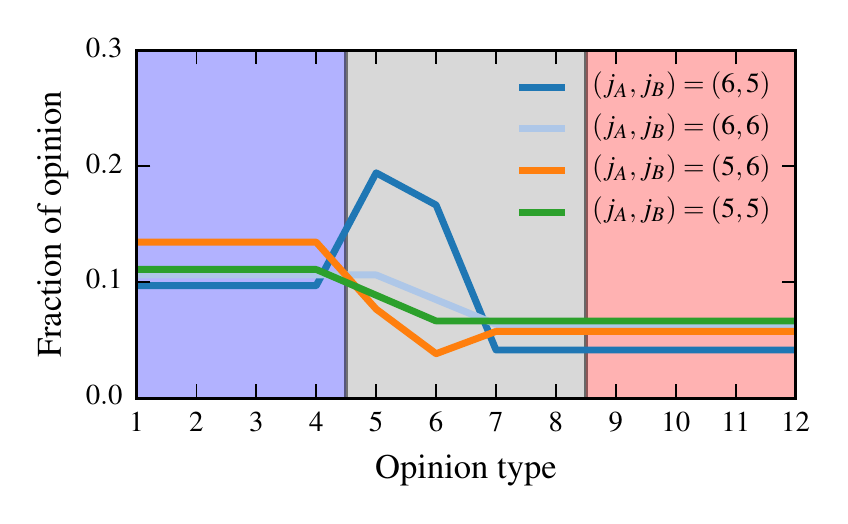}
	\end{minipage}
		\begin{minipage}{0.49\textwidth}
		\centering
		\includegraphics[scale=.9]{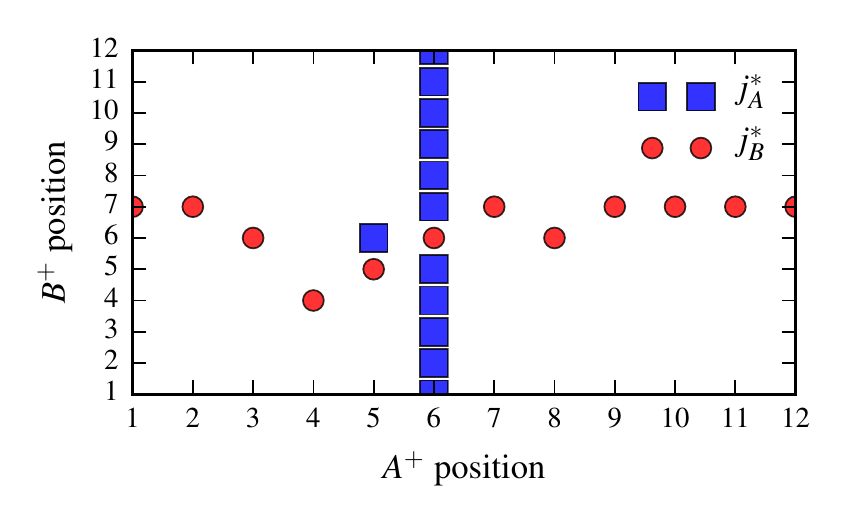}
	\end{minipage}
	  \caption{
	  	\textbf{Stationary distributions and best responses for an even number of states.}
	  	Based on Eq.~(6) in the main text, the stationary solution is computed for two activists at different locations which lead to locally larger ($p+\epsilon_A$, $p+\epsilon_B$) or smaller ($p-\epsilon_A$, $p-\epsilon_B$) transition probabilities ($p=0.2$). The number of states is $N=12$. From the upper to the lower panels, we assumed different values of $\epsilon_A \in\{0.1,0.11,0.15\}$ and set $\epsilon_B=0.1$. All states left from state 7 belong to group $A$, whereas all states right from state $6$ belong to group $B$. The left panels show the stationary distributions for different locations of activists $j_A$ and $j_B$. The right panels are the corresponding best response curves. There exist four Nash equilibria for $\epsilon_A=\epsilon_B$ and dynamic equilibria otherwise.
  	 } 
	 \label{fig:even}
\end{figure}
In the main part of the manuscript, we described the emergence of dynamic equilibria for $\epsilon_A>\epsilon_A^c$. However, we only focused on an odd number of states ($N=9$). In the case of an even number states, there exists no unique center. Due to the missing center, there are also multiple Nash equilibria, as shown in the upper panels of Fig.~\ref{fig:even} where we set $\epsilon_A=\epsilon_B=0.1$ and $N=12$. Interestingly, two equilibria correspond to uniform stationary distributions, whereas the remaining ones describe a polarized and an unpolarized population. Unlike in the case of a clearly defined center, we find the possibility of an emerging polarization or anti-polarization effect for equally strong activists. A slight increase of the value of $\epsilon_A$ leads to the disappearance of the Nash equilibria. Instead we find a dynamic equilibrium where activist $A^+$ avoids its opponent and moves to the right first. Increasing the value of $\epsilon_A$ even more makes activist $A^+$ avoid its opponent by moving to the left. This behavior is in accordance with the observations made in the main text.
\end{document}